\documentclass[zpreprint,zbstdefault,british]{./zeus_K2.9paper}
\usepackage{./zeusdet_units}
\chardef\usc=95
\chardef\til=126
\catcode`\@=11 
\DeclareRobustCommand\xdotspace{\futurelet\@let@token\@xdotspace}
\def\@xdotspace{%
  \ifx\@let@token.\else
  \ifx\@let@token\bgroup.\else
  \ifx\@let@token\egroup.\else
  \ifx\@let@token\/.\else
  \ifx\@let@token\ .\else
  \ifx\@let@token~.\else
  \ifx\@let@token!.\else
  \ifx\@let@token,.\else
  \ifx\@let@token:.\else
  \ifx\@let@token;.\else
  \ifx\@let@token?.\else
  \ifx\@let@token/.\else
  \ifx\@let@token'.\else
  \ifx\@let@token).\else
  \ifx\@let@token-.\else
  \ifx\@let@token\@xobeysp.\else
  \ifx\@let@token\space.\else
  \ifx\@let@token\@sptoken.\else
   .\space
   \fi\fi\fi\fi\fi\fi\fi\fi\fi\fi\fi\fi\fi\fi\fi\fi\fi\fi}
\catcode`\@=12 

\newcommand{\stru}[2]{%
   \relax\ifmmode\hbox{\vrule height#1 depth#2 width0pt}%
   \else\vrule height#1 depth#2 width0pt\fi}

\newcommand{\Ronum}[1]{\uppercase\expandafter{\romannumeral#1}}
\newcommand{\ronum}[1]{\expandafter{\romannumeral#1}}
\DeclareRobustCommand{\LaTeXZ}{%
  \LaTeX\kern-.05em4\kern-.1em
  {\raisebox{-0.2ex}{$\scriptstyle\text{ZEUS}$}}\xspace}

\newcommand{\slashfrac}[2]{%
  \raisebox{0.5ex}{\ensuremath #1}\kern-0.12em/\kern-0.08em
  \raisebox{-.8ex}{\ensuremath #2}}

\newcommand{\sqr}[3]{%
    {\vcenter{\hrule height.#3ex\hbox{\vrule width.#2ex height#1ex
     \kern#1ex\vrule width.#3ex}\hrule height.#2ex}}}

\catcode`\@=11 
\newcommand{\parenbar}{\mathpalette\p@renb@r}
\def\p@renb@r#1#2{\vbox{%
  \ifx#1\scriptscriptstyle \dimen@.7em\dimen@ii.2em\else
  \ifx#1\scriptstyle \dimen@.8em\dimen@ii.25em\else
  \dimen@1em\dimen@ii.4em\fi\fi \offinterlineskip
  \ialign{\hfill##\hfill\cr
    \vbox{\hrule width\dimen@ii}\cr
    \noalign{\vskip-.3ex}%
    \hbox to\dimen@{$\mathchar300\hfil\mathchar301$}\cr
    \noalign{\vskip-.3ex}%
    $#1#2$\cr}}}
\catcode`\@=12 

\ifzeusunit
  
\else
  
\fi

\newcommand{\CC}{{\rm CC}}

\newcommand{\DA}{{\rm DA}}

\newcommand{\IP}{{\rm I$\kern-0.01667em$P}\xspace}
\newcommand{\JB}{{\rm JB}}

\newcommand{\NC}{{\rm NC}}

\mathchardef\qsm=63
\mathchardef\pls=43
\mathchardef\mns=512
\mathchardef\plm=518
\mathchardef\eql=61
\mathchardef\smallleft=300
\mathchardef\smallright=301
\mathchardef\les=316
\mathchardef\gre=318
\mathchardef\leq=532
\mathchardef\grq=533

\catcode`\@=11 
\newcounter{pict@width}
\newcounter{pict@height}
\newlength{\pict@scale}
\newcommand{\psfigadd}[4]{%
\setcounter{pict@width}{1*\ratio{#2+\pict@scale/2}{\pict@scale}}
\setcounter{pict@height}{1*\ratio{#3+\pict@scale/2}{\pict@scale}}
\setlength{\unitlength}{\pict@scale}
\hbox to #2{\hspace{-\fill}\begin{picture}(\thepict@width,\thepict@height)
\put(0,0){\psfig{figure=#1,width=#2,height=#3,clip=}}
\SetScale{0.283466457}
\SetWidth{1.763889}
{#4}
\end{picture}}
}
\newcounter{pict@widthfst}
\newcounter{pict@widthscd}
\newcounter{pict@widthtot}
\newcommand{\psfigaddtwo}[7]{%
\setcounter{pict@widthfst}{1*\ratio{#2+\pict@scale/2}{\pict@scale}}
\setcounter{pict@widthscd}{1*\ratio{#2+#4+\pict@scale/2}{\pict@scale}}
\setcounter{pict@widthtot}{1*\ratio{#2+#4+#6+\pict@scale/2}{\pict@scale}}
\setcounter{pict@height}{1*\ratio{#3+\pict@scale/2}{\pict@scale}}
\setlength{\unitlength}{\pict@scale}
\hbox{\hspace{-\fill}\begin{picture}(\thepict@widthtot,\thepict@height)
\put(0,0){\psfig{figure=#1,width=#2,height=#3,clip=}}
\put(\thepict@widthscd,0){\psfig{figure=#5,width=#6,height=#3,clip=}}
\SetScale{0.283466457}
\SetWidth{1.763889}
{#7}
\end{picture}}
}
\newcommand{\psfigror}[4]{%
\setcounter{pict@width}{1*\ratio{#2+\pict@scale/2}{\pict@scale}}
\setcounter{pict@height}{1*\ratio{#3+\pict@scale/2}{\pict@scale}}
\setlength{\unitlength}{\pict@scale}
\hbox{\begin{picture}(\thepict@width,\thepict@height)
\put(0,\thepict@height){\psfig{figure=#1,width=#3,height=#2,clip=,angle=270}}
\SetScale{0.283466457}
\SetWidth{1.763889}
{#4}
\end{picture}}
}
\newcommand{\psfigrol}[4]{%
\setcounter{pict@width}{1*\ratio{#2+\pict@scale/2}{\pict@scale}}
\setcounter{pict@height}{1*\ratio{#3+\pict@scale/2}{\pict@scale}}
\setlength{\unitlength}{\pict@scale}
\hbox{\begin{picture}(\thepict@width,\thepict@height)
\put(0,0){\psfig{figure=#1,width=#3,height=#2,clip=,angle=90}}
\SetScale{0.283466457}
\SetWidth{1.763889}
{#4}
\end{picture}}
}
\catcode`\@=12 
\newlength\listtextwidth

\catcode`\@=11 
\newlength{\@tabfninsert}
\newlength{\@tabfnwidth}
\newcommand{\tabfootnote}[2]{%
  \setlength{\@tabfninsert}{0.8em}
  \setlength{\@tabfnwidth}{\textwidth}
  \addtolength{\@tabfnwidth}{-\@tabfninsert}
  \addtolength{\@tabfnwidth}{-0.4em}
  \noindent\makebox[\@tabfninsert][r]{\footnotesize$^{#1}$\hfil}\hfill%
  \parbox[t]{\@tabfnwidth}{\footnotesize #2\hfill}}
\catcode`\@=12 

\def\citeCTD{{\cite{%
nim:a279:290,*npps:b32:181,*nim:a338:254%
}}\xspace}
\def\citeMVD{{\cite{%
nim:a581:656%
}}\xspace}
\def\citeSTT{{\cite{%
nim:a535:191%
}}\xspace}
\def\citeCAL{{\cite{%
nim:a309:77,*nim:a309:101,*nim:a321:356,*nim:a336:23%
}}\xspace}
\def\citeHES{{\cite{%
nim:a277:176%
}}\xspace}
\def\citeSixmT{{\cite{%
thesis:gosau:2007%
}}\xspace}
\def\citeBAC{{\cite{%
nim:a300:480%
}}\xspace}
\def\citePCAL{{\cite{%
desy-92-066,*zfp:c63:391,*acpp:b32:2025%
}}\xspace}
\def\citeSPECTRO{{\cite{%
nim:a565:572%
}}\xspace}

\begin{document}
%
%
\prepnum{DESY--12--077}
\prepdate{23 May 2012}

\zeustitle{%
  Search for first-generation leptoquarks at HERA
}
                    
\zeusauthor{ZEUS Collaboration}
\zeusdate{}

\maketitle

%
%
\begin{abstract}\noindent
A search for first-generation leptoquarks was performed in electron-proton and positron-proton collisions recorded with the ZEUS detector at HERA in 2003--2007 using an integrated luminosity of \unit[366]{$\pbi$}. Final states with an electron and jets or with missing transverse momentum and jets were analysed, searching for resonances or other deviations from the Standard Model predictions. No evidence for any leptoquark signal was found. The data were combined with data previously taken at HERA, resulting in a total integrated luminosity of \unit[498]{$\pbi$}. Limits on the Yukawa coupling, $\lambda$, of leptoquarks were set as a function of the leptoquark mass for different leptoquark types within the Buchm\"uller-R\"uckl-Wyler model. Leptoquarks with a coupling $\lambda=0.3$ are excluded for masses up to \unit[699]{\Gev}.
\end{abstract}

\thispagestyle{empty}
\clearpage
%
%
\pagenumbering{Roman} 

\begin{center}
{                      \Large  The ZEUS Collaboration              }
\end{center}

{\small


        {\raggedright
H.~Abramowicz$^{45, aj}$, 
I.~Abt$^{35}$, 
L.~Adamczyk$^{13}$, 
M.~Adamus$^{54}$, 
R.~Aggarwal$^{7, c}$, 
S.~Antonelli$^{4}$, 
P.~Antonioli$^{3}$, 
A.~Antonov$^{33}$, 
M.~Arneodo$^{50}$, 
O.~Arslan$^{5}$
% ...
        }

\newpage


\makebox[3em]{$^{1}$}
\begin{minipage}[t]{14cm}
{\it Argonne National Laboratory, Argonne, Illinois 60439-4815, USA}~$^{A}$

\end{minipage}\\
\makebox[3em]{$^{2}$}
\begin{minipage}[t]{14cm}
{\it Andrews University, Berrien Springs, Michigan 49104-0380, USA}

\end{minipage}\\
\makebox[3em]{$^{3}$}
\begin{minipage}[t]{14cm}
{\it INFN Bologna, Bologna, Italy}~$^{B}$

\end{minipage}\\
\makebox[3em]{$^{4}$}
\begin{minipage}[t]{14cm}
{\it University and INFN Bologna, Bologna, Italy}~$^{B}$

\end{minipage}\\
\makebox[3em]{$^{5}$}
\begin{minipage}[t]{14cm}
{\it Physikalisches Institut der Universit\"at Bonn,
Bonn, Germany}~$^{C}$

\end{minipage}\\
\makebox[3em]{$^{6}$}
\begin{minipage}[t]{14cm}
{\it H.H.~Wills Physics Laboratory, University of Bristol,
Bristol, United Kingdom}~$^{D}$

\end{minipage}\\
\makebox[3em]{$^{7}$}
\begin{minipage}[t]{14cm}
{\it Panjab University, Department of Physics, Chandigarh, India}

\end{minipage}\\
\makebox[3em]{$^{8}$}
\begin{minipage}[t]{14cm}
{\it Calabria University,
Physics Department and INFN, Cosenza, Italy}~$^{B}$

\end{minipage}\\
\makebox[3em]{$^{9}$}
\begin{minipage}[t]{14cm}
{\it Institute for Universe and Elementary Particles, Chonnam National University,\\
Kwangju, South Korea}

\end{minipage}\\
\makebox[3em]{$^{10}$}
\begin{minipage}[t]{14cm}
{\it Jabatan Fizik, Universiti Malaya, 50603 Kuala Lumpur, Malaysia}~$^{E}$

\end{minipage}\\
\makebox[3em]{$^{11}$}
\begin{minipage}[t]{14cm}
{\it Nevis Laboratories, Columbia University, Irvington on Hudson,
New York 10027, USA}~$^{F}$

\end{minipage}\\
\makebox[3em]{$^{12}$}
\begin{minipage}[t]{14cm}
{\it The Henryk Niewodniczanski Institute of Nuclear Physics, Polish Academy of \\
Sciences, Krakow, Poland}~$^{G}$

\end{minipage}\\
\makebox[3em]{$^{13}$}
\begin{minipage}[t]{14cm}
{\it AGH-University of Science and Technology, Faculty of Physics and Applied Computer
Science, Krakow, Poland}~$^{H}$

\end{minipage}\\
\makebox[3em]{$^{14}$}
\begin{minipage}[t]{14cm}
{\it Department of Physics, Jagellonian University, Cracow, Poland}

\end{minipage}\\
\makebox[3em]{$^{15}$}
\begin{minipage}[t]{14cm}
{\it Deutsches Elektronen-Synchrotron DESY, Hamburg, Germany}

\end{minipage}\\
\makebox[3em]{$^{16}$}
\begin{minipage}[t]{14cm}
{\it Deutsches Elektronen-Synchrotron DESY, Zeuthen, Germany}

\end{minipage}\\
\makebox[3em]{$^{17}$}
\begin{minipage}[t]{14cm}
{\it INFN Florence, Florence, Italy}~$^{B}$

\end{minipage}\\
\makebox[3em]{$^{18}$}
\begin{minipage}[t]{14cm}
{\it University and INFN Florence, Florence, Italy}~$^{B}$

\end{minipage}\\
\makebox[3em]{$^{19}$}
\begin{minipage}[t]{14cm}
{\it Fakult\"at f\"ur Physik der Universit\"at Freiburg i.Br.,
Freiburg i.Br., Germany}

\end{minipage}\\
\makebox[3em]{$^{20}$}
\begin{minipage}[t]{14cm}
{\it School of Physics and Astronomy, University of Glasgow,
Glasgow, United Kingdom}~$^{D}$

\end{minipage}\\
\makebox[3em]{$^{21}$}
\begin{minipage}[t]{14cm}
{\it Department of Engineering in Management and Finance, Univ. of
the Aegean, Chios, Greece}

\end{minipage}\\
\makebox[3em]{$^{22}$}
\begin{minipage}[t]{14cm}
{\it Hamburg University, Institute of Experimental Physics, Hamburg,
Germany}~$^{I}$

\end{minipage}\\
\makebox[3em]{$^{23}$}
\begin{minipage}[t]{14cm}
{\it Imperial College London, High Energy Nuclear Physics Group,
London, United Kingdom}~$^{D}$

\end{minipage}\\
\makebox[3em]{$^{24}$}
\begin{minipage}[t]{14cm}
{\it Institute of Particle and Nuclear Studies, KEK,
Tsukuba, Japan}~$^{J}$

\end{minipage}\\
\makebox[3em]{$^{25}$}
\begin{minipage}[t]{14cm}
{\it Institute of Physics and Technology of Ministry of Education and
Science of Kazakhstan, Almaty, Kazakhstan}

\end{minipage}\\
\makebox[3em]{$^{26}$}
\begin{minipage}[t]{14cm}
{\it Institute for Nuclear Research, National Academy of Sciences, Kyiv, Ukraine}

\end{minipage}\\
\makebox[3em]{$^{27}$}
\begin{minipage}[t]{14cm}
{\it Department of Nuclear Physics, National Taras Shevchenko University of Kyiv, Kyiv, Ukraine}

\end{minipage}\\
\makebox[3em]{$^{28}$}
\begin{minipage}[t]{14cm}
{\it Kyungpook National University, Center for High Energy Physics, Daegu,
South Korea}~$^{K}$

\end{minipage}\\
\makebox[3em]{$^{29}$}
\begin{minipage}[t]{14cm}
{\it Institut de Physique Nucl\'{e}aire, Universit\'{e} Catholique de Louvain, Louvain-la-Neuve,\\
Belgium}~$^{L}$

\end{minipage}\\
\makebox[3em]{$^{30}$}
\begin{minipage}[t]{14cm}
{\it Departamento de F\'{\i}sica Te\'orica, Universidad Aut\'onoma
de Madrid, Madrid, Spain}~$^{M}$

\end{minipage}\\
\makebox[3em]{$^{31}$}
\begin{minipage}[t]{14cm}
{\it Department of Physics, McGill University,
Montr\'eal, Qu\'ebec, Canada H3A 2T8}~$^{N}$

\end{minipage}\\
\makebox[3em]{$^{32}$}
\begin{minipage}[t]{14cm}
{\it Meiji Gakuin University, Faculty of General Education,
Yokohama, Japan}~$^{J}$

\end{minipage}\\
\makebox[3em]{$^{33}$}
\begin{minipage}[t]{14cm}
{\it Moscow Engineering Physics Institute, Moscow, Russia}~$^{O}$

\end{minipage}\\
\makebox[3em]{$^{34}$}
\begin{minipage}[t]{14cm}
{\it Lomonosov Moscow State University, Skobeltsyn Institute of Nuclear Physics,
Moscow, Russia}~$^{P}$

\end{minipage}\\
\makebox[3em]{$^{35}$}
\begin{minipage}[t]{14cm}
{\it Max-Planck-Institut f\"ur Physik, M\"unchen, Germany}

\end{minipage}\\
\makebox[3em]{$^{36}$}
\begin{minipage}[t]{14cm}
{\it NIKHEF and University of Amsterdam, Amsterdam, Netherlands}~$^{Q}$

\end{minipage}\\
\makebox[3em]{$^{37}$}
\begin{minipage}[t]{14cm}
{\it Physics Department, Ohio State University,
Columbus, Ohio 43210, USA}~$^{A}$

\end{minipage}\\
\makebox[3em]{$^{38}$}
\begin{minipage}[t]{14cm}
{\it Department of Physics, University of Oxford,
Oxford, United Kingdom}~$^{D}$

\end{minipage}\\
\makebox[3em]{$^{39}$}
\begin{minipage}[t]{14cm}
{\it INFN Padova, Padova, Italy}~$^{B}$

\end{minipage}\\
\makebox[3em]{$^{40}$}
\begin{minipage}[t]{14cm}
{\it Dipartimento di Fisica dell' Universit\`a and INFN,
Padova, Italy}~$^{B}$

\end{minipage}\\
\makebox[3em]{$^{41}$}
\begin{minipage}[t]{14cm}
{\it Department of Physics, Pennsylvania State University, University Park,\\
Pennsylvania 16802, USA}~$^{F}$

\end{minipage}\\
\makebox[3em]{$^{42}$}
\begin{minipage}[t]{14cm}
{\it Polytechnic University, Tokyo, Japan}~$^{J}$

\end{minipage}\\
\makebox[3em]{$^{43}$}
\begin{minipage}[t]{14cm}
{\it Dipartimento di Fisica, Universit\`a 'La Sapienza' and INFN,
Rome, Italy}~$^{B}$

\end{minipage}\\
\makebox[3em]{$^{44}$}
\begin{minipage}[t]{14cm}
{\it Rutherford Appleton Laboratory, Chilton, Didcot, Oxon,
United Kingdom}~$^{D}$

\end{minipage}\\
\makebox[3em]{$^{45}$}
\begin{minipage}[t]{14cm}
{\it Raymond and Beverly Sackler Faculty of Exact Sciences, School of Physics, \\
Tel Aviv University, Tel Aviv, Israel}~$^{R}$

\end{minipage}\\
\makebox[3em]{$^{46}$}
\begin{minipage}[t]{14cm}
{\it Department of Physics, Tokyo Institute of Technology,
Tokyo, Japan}~$^{J}$

\end{minipage}\\
\makebox[3em]{$^{47}$}
\begin{minipage}[t]{14cm}
{\it Department of Physics, University of Tokyo,
Tokyo, Japan}~$^{J}$

\end{minipage}\\
\makebox[3em]{$^{48}$}
\begin{minipage}[t]{14cm}
{\it Tokyo Metropolitan University, Department of Physics,
Tokyo, Japan}~$^{J}$

\end{minipage}\\
\makebox[3em]{$^{49}$}
\begin{minipage}[t]{14cm}
{\it Universit\`a di Torino and INFN, Torino, Italy}~$^{B}$

\end{minipage}\\
\makebox[3em]{$^{50}$}
\begin{minipage}[t]{14cm}
{\it Universit\`a del Piemonte Orientale, Novara, and INFN, Torino,
Italy}~$^{B}$

\end{minipage}\\
\makebox[3em]{$^{51}$}
\begin{minipage}[t]{14cm}
{\it Department of Physics, University of Toronto, Toronto, Ontario,
Canada M5S 1A7}~$^{N}$

\end{minipage}\\
\makebox[3em]{$^{52}$}
\begin{minipage}[t]{14cm}
{\it Physics and Astronomy Department, University College London,
London, United Kingdom}~$^{D}$

\end{minipage}\\
\makebox[3em]{$^{53}$}
\begin{minipage}[t]{14cm}
{\it Faculty of Physics, University of Warsaw, Warsaw, Poland}

\end{minipage}\\
\makebox[3em]{$^{54}$}
\begin{minipage}[t]{14cm}
{\it National Centre for Nuclear Research, Warsaw, Poland}

\end{minipage}\\
\makebox[3em]{$^{55}$}
\begin{minipage}[t]{14cm}
{\it Department of Particle Physics and Astrophysics, Weizmann
Institute, Rehovot, Israel}

\end{minipage}\\
\makebox[3em]{$^{56}$}
\begin{minipage}[t]{14cm}
{\it Department of Physics, University of Wisconsin, Madison,
Wisconsin 53706, USA}~$^{A}$

\end{minipage}\\
\makebox[3em]{$^{57}$}
\begin{minipage}[t]{14cm}
{\it Department of Physics, York University, Ontario, Canada M3J 1P3}~$^{N}$

\end{minipage}\\
\vspace{30em} \pagebreak[4]


\makebox[3ex]{$^{ A}$}
\begin{minipage}[t]{14cm}
 supported by the US Department of Energy\
\end{minipage}\\
\makebox[3ex]{$^{ B}$}
\begin{minipage}[t]{14cm}
 supported by the Italian National Institute for Nuclear Physics (INFN) \
\end{minipage}\\
\makebox[3ex]{$^{ C}$}
\begin{minipage}[t]{14cm}
 supported by the German Federal Ministry for Education and Research (BMBF), under
 contract No. 05 H09PDF\
\end{minipage}\\
\makebox[3ex]{$^{ D}$}
\begin{minipage}[t]{14cm}
 supported by the Science and Technology Facilities Council, UK\
\end{minipage}\\
\makebox[3ex]{$^{ E}$}
\begin{minipage}[t]{14cm}
 supported by an FRGS grant from the Malaysian government\
\end{minipage}\\
\makebox[3ex]{$^{ F}$}
\begin{minipage}[t]{14cm}
 supported by the US National Science Foundation. Any opinion,
 findings and conclusions or recommendations expressed in this material
 are those of the authors and do not necessarily reflect the views of the
 National Science Foundation.\
\end{minipage}\\
\makebox[3ex]{$^{ G}$}
\begin{minipage}[t]{14cm}
 supported by the Polish Ministry of Science and Higher Education as a scientific project No.
 DPN/N188/DESY/2009\
\end{minipage}\\
\makebox[3ex]{$^{ H}$}
\begin{minipage}[t]{14cm}
 supported by the Polish Ministry of Science and Higher Education and its grants
 for Scientific Research\
\end{minipage}\\
\makebox[3ex]{$^{ I}$}
\begin{minipage}[t]{14cm}
 supported by the German Federal Ministry for Education and Research (BMBF), under
 contract No. 05h09GUF, and the SFB 676 of the Deutsche Forschungsgemeinschaft (DFG) \
\end{minipage}\\
\makebox[3ex]{$^{ J}$}
\begin{minipage}[t]{14cm}
 supported by the Japanese Ministry of Education, Culture, Sports, Science and Technology
 (MEXT) and its grants for Scientific Research\
\end{minipage}\\
\makebox[3ex]{$^{ K}$}
\begin{minipage}[t]{14cm}
 supported by the Korean Ministry of Education and Korea Science and Engineering
 Foundation\
\end{minipage}\\
\makebox[3ex]{$^{ L}$}
\begin{minipage}[t]{14cm}
 supported by FNRS and its associated funds (IISN and FRIA) and by an Inter-University
 Attraction Poles Programme subsidised by the Belgian Federal Science Policy Office\
\end{minipage}\\
\makebox[3ex]{$^{ M}$}
\begin{minipage}[t]{14cm}
 supported by the Spanish Ministry of Education and Science through funds provided by
 CICYT\
\end{minipage}\\
\makebox[3ex]{$^{ N}$}
\begin{minipage}[t]{14cm}
 supported by the Natural Sciences and Engineering Research Council of Canada (NSERC) \
\end{minipage}\\
\makebox[3ex]{$^{ O}$}
\begin{minipage}[t]{14cm}
 partially supported by the German Federal Ministry for Education and Research (BMBF)\
\end{minipage}\\
\makebox[3ex]{$^{ P}$}
\begin{minipage}[t]{14cm}
 supported by RF Presidential grant N 4142.2010.2 for Leading Scientific Schools, by the
 Russian Ministry of Education and Science through its grant for Scientific Research on
 High Energy Physics and under contract No.02.740.11.0244 \
\end{minipage}\\
\makebox[3ex]{$^{ Q}$}
\begin{minipage}[t]{14cm}
 supported by the Netherlands Foundation for Research on Matter (FOM)\
\end{minipage}\\
\makebox[3ex]{$^{ R}$}
\begin{minipage}[t]{14cm}
 supported by the Israel Science Foundation\
\end{minipage}\\
\vspace{30em} \pagebreak[4]


\makebox[3ex]{$^{ a}$}
\begin{minipage}[t]{14cm}
now at University of Salerno, Italy\
\end{minipage}\\
\makebox[3ex]{$^{ b}$}
\begin{minipage}[t]{14cm}
now at Queen Mary University of London, United Kingdom\
\end{minipage}\\
\makebox[3ex]{$^{ c}$}
\begin{minipage}[t]{14cm}
also funded by Max Planck Institute for Physics, Munich, Germany\
\end{minipage}\\
\makebox[3ex]{$^{ d}$}
\begin{minipage}[t]{14cm}
also Senior Alexander von Humboldt Research Fellow at Hamburg University,
 Institute of Experimental Physics, Hamburg, Germany\
\end{minipage}\\
\makebox[3ex]{$^{ e}$}
\begin{minipage}[t]{14cm}
also at Cracow University of Technology, Faculty of Physics,
 Mathemathics and Applied Computer Science, Poland\
\end{minipage}\\
\makebox[3ex]{$^{ f}$}
\begin{minipage}[t]{14cm}
supported by the research grant No. 1 P03B 04529 (2005-2008)\
\end{minipage}\\
\makebox[3ex]{$^{ g}$}
\begin{minipage}[t]{14cm}
supported by the Polish National Science Centre, project No. DEC-2011/01/BST2/03643\
\end{minipage}\\
\makebox[3ex]{$^{ h}$}
\begin{minipage}[t]{14cm}
now at Rockefeller University, New York, NY
 10065, USA\
\end{minipage}\\
\makebox[3ex]{$^{ i}$}
\begin{minipage}[t]{14cm}
now at DESY group FS-CFEL-1\
\end{minipage}\\
\makebox[3ex]{$^{ j}$}
\begin{minipage}[t]{14cm}
now at Institute of High Energy Physics, Beijing, China\
\end{minipage}\\
\makebox[3ex]{$^{ k}$}
\begin{minipage}[t]{14cm}
now at DESY group FEB, Hamburg, Germany\
\end{minipage}\\
\makebox[3ex]{$^{ l}$}
\begin{minipage}[t]{14cm}
also at Moscow State University, Russia\
\end{minipage}\\
\makebox[3ex]{$^{ m}$}
\begin{minipage}[t]{14cm}
now at University of Liverpool, United Kingdom\
\end{minipage}\\
\makebox[3ex]{$^{ n}$}
\begin{minipage}[t]{14cm}
now at CERN, Geneva, Switzerland\
\end{minipage}\\
\makebox[3ex]{$^{ o}$}
\begin{minipage}[t]{14cm}
also affiliated with Universtiy College London, UK\
\end{minipage}\\
\makebox[3ex]{$^{ p}$}
\begin{minipage}[t]{14cm}
now at Goldman Sachs, London, UK\
\end{minipage}\\
\makebox[3ex]{$^{ q}$}
\begin{minipage}[t]{14cm}
also at Institute of Theoretical and Experimental Physics, Moscow, Russia\
\end{minipage}\\
\makebox[3ex]{$^{ r}$}
\begin{minipage}[t]{14cm}
also at FPACS, AGH-UST, Cracow, Poland\
\end{minipage}\\
\makebox[3ex]{$^{ s}$}
\begin{minipage}[t]{14cm}
partially supported by Warsaw University, Poland\
\end{minipage}\\
\makebox[3ex]{$^{ t}$}
\begin{minipage}[t]{14cm}
now at Istituto Nucleare di Fisica Nazionale (INFN), Pisa, Italy\
\end{minipage}\\
\makebox[3ex]{$^{ u}$}
\begin{minipage}[t]{14cm}
now at Haase Energie Technik AG, Neum\"unster, Germany\
\end{minipage}\\
\makebox[3ex]{$^{ v}$}
\begin{minipage}[t]{14cm}
now at Department of Physics, University of Bonn, Germany\
\end{minipage}\\
\makebox[3ex]{$^{ w}$}
\begin{minipage}[t]{14cm}
now at Biodiversit\"at und Klimaforschungszentrum (BiK-F), Frankfurt, Germany\
\end{minipage}\\
\makebox[3ex]{$^{ x}$}
\begin{minipage}[t]{14cm}
also affiliated with DESY, Germany\
\end{minipage}\\
\makebox[3ex]{$^{ y}$}
\begin{minipage}[t]{14cm}
also at University of Tokyo, Japan\
\end{minipage}\\
\makebox[3ex]{$^{ z}$}
\begin{minipage}[t]{14cm}
now at Kobe University, Japan\
\end{minipage}\\
\makebox[3ex]{$^{\dagger}$}
\begin{minipage}[t]{14cm}
 deceased \
\end{minipage}\\
\makebox[3ex]{$^{aa}$}
\begin{minipage}[t]{14cm}
supported by DESY, Germany\
\end{minipage}\\
\makebox[3ex]{$^{ab}$}
\begin{minipage}[t]{14cm}
member of National Technical University of Ukraine, Kyiv Polytechnic Institute,
 Kyiv, Ukraine\
\end{minipage}\\
\makebox[3ex]{$^{ac}$}
\begin{minipage}[t]{14cm}
member of National University of Kyiv - Mohyla Academy, Kyiv, Ukraine\
\end{minipage}\\
\makebox[3ex]{$^{ad}$}
\begin{minipage}[t]{14cm}
partly supported by the Russian Foundation for Basic Research, grant 11-02-91345-DFG\_a\
\end{minipage}\\
\makebox[3ex]{$^{ae}$}
\begin{minipage}[t]{14cm}
Alexander von Humboldt Professor; also at DESY and University of Oxford\
\end{minipage}\\
\makebox[3ex]{$^{af}$}
\begin{minipage}[t]{14cm}
STFC Advanced Fellow\
\end{minipage}\\
\makebox[3ex]{$^{ag}$}
\begin{minipage}[t]{14cm}
nee Korcsak-Gorzo\
\end{minipage}\\
\makebox[3ex]{$^{ah}$}
\begin{minipage}[t]{14cm}
now at LNF, Frascati, Italy\
\end{minipage}\\
\makebox[3ex]{$^{ai}$}
\begin{minipage}[t]{14cm}
This material was based on work supported by the
 National Science Foundation, while working at the Foundation.\
\end{minipage}\\
\makebox[3ex]{$^{aj}$}
\begin{minipage}[t]{14cm}
also at Max Planck Institute for Physics, Munich, Germany, External Scientific Member\
\end{minipage}\\
\makebox[3ex]{$^{ak}$}
\begin{minipage}[t]{14cm}
now at Tokyo Metropolitan University, Japan\
\end{minipage}\\
\makebox[3ex]{$^{al}$}
\begin{minipage}[t]{14cm}
now at Nihon Institute of Medical Science, Japan\
\end{minipage}\\
\makebox[3ex]{$^{am}$}
\begin{minipage}[t]{14cm}
now at Osaka University, Osaka, Japan\
\end{minipage}\\
\makebox[3ex]{$^{an}$}
\begin{minipage}[t]{14cm}
also at \L\'{o}d\'{z} University, Poland\
\end{minipage}\\
\makebox[3ex]{$^{ao}$}
\begin{minipage}[t]{14cm}
member of \L\'{o}d\'{z} University, Poland\
\end{minipage}\\
\makebox[3ex]{$^{ap}$}
\begin{minipage}[t]{14cm}
now at Department of Physics, Stockholm University, Stockholm, Sweden\
\end{minipage}\\
\makebox[3ex]{$^{aq}$}
\begin{minipage}[t]{14cm}
also at Cardinal Stefan Wyszy\'nski University, Warsaw, Poland\
\end{minipage}\\

}
\clearpage
\pagenumbering{arabic}
%
%
\section{Introduction}
\label{sec-int}

\begin{figure}[p]
\vfill
\begin{center}
\includegraphics[width=16cm]{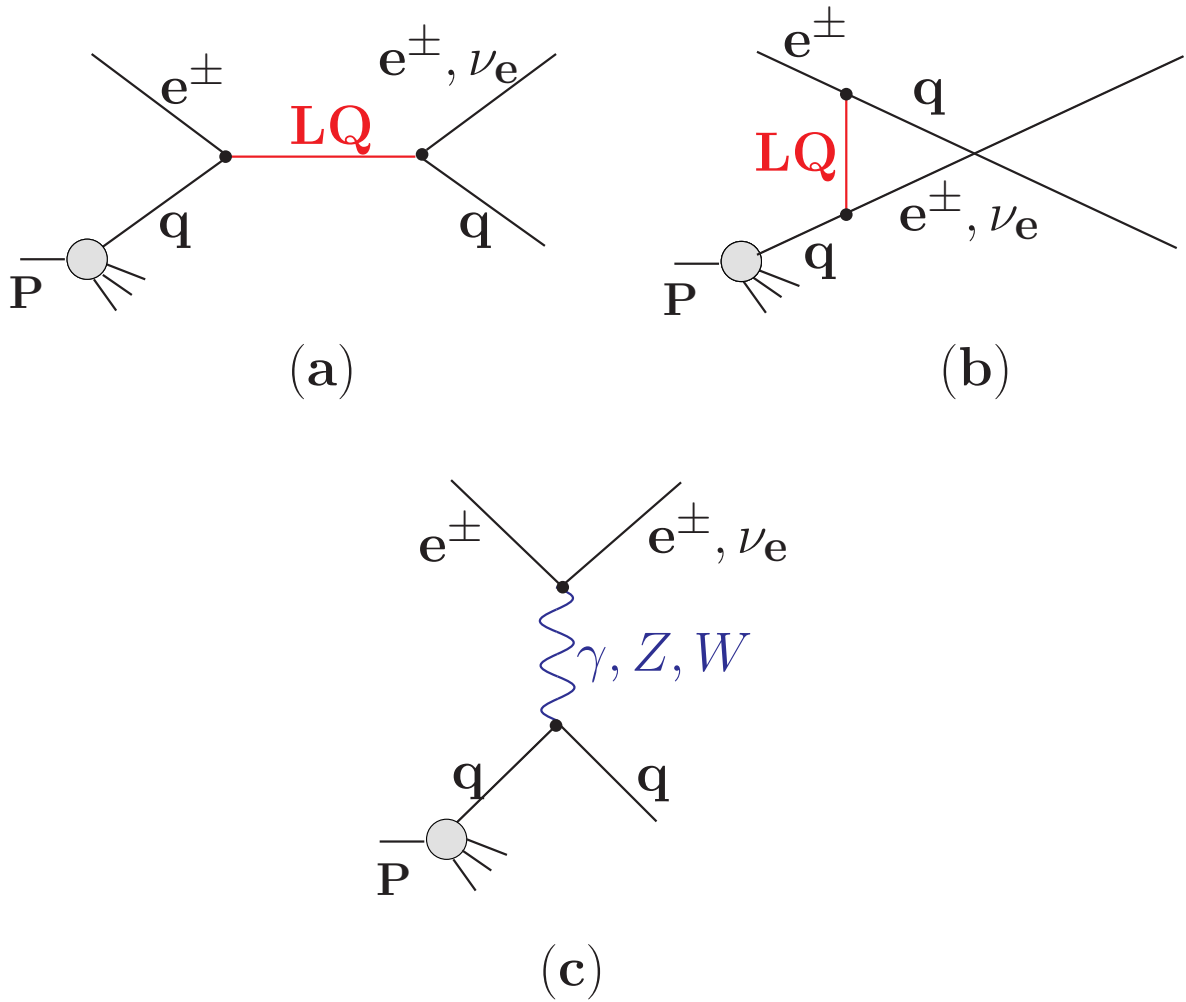}
\end{center}
\caption{Diagrams for 
(a) $s$-channel
LQ production/exchange and (b) $u$-channel LQ exchange and 
for (c) SM deep inelastic scattering  
via photon, $Z^0$ and $W$ exchange.}
\label{fig-feynman}
\vfill
\end{figure}

Many extensions of the Standard Model (SM)
predict the existence of
particles carrying both baryon and lepton number, such as
leptoquarks (LQs)~\cite{pl:b191:442}. 
In $ep$ collisions at HERA, such states could have been produced 
directly through electron\footnote{Unless otherwise specified,
`electron' refers to both positron and electron and `neutrino' 
refers to both neutrino 
and antineutrino.}-quark fusion (Fig.~\ref{fig-feynman}a) 
if their masses, $M_\mathrm{LQ}$, were lower than the 
HERA centre-of-mass energy, $\sqrt{s}$. 
The leptoquarks would have decayed into  
an electron and quark or an electron neutrino and quark,
yielding peaks in the spectra of
the final-state lepton-jets\footnote{There can be more than one 
jet in the final state due to QCD initial or final state radiation.} 
invariant mass, $M_{ljs}$. 
Leptoquarks with $M_\mathrm{LQ} > \sqrt{s}$ could not have been produced 
as resonances, but they would still have caused deviations from the 
SM prediction in the observed $M_{ljs}$ spectrum due to virtual LQ exchange
(Fig.~\ref{fig-feynman}a and b). 
This paper presents an analysis of the $M_{ljs}$ spectrum 
searching for evidence for leptoquarks.  

The prediction for the $M_{ljs}$ spectrum is given by SM neutral current (NC) 
and charged current (CC) deep inelastic scattering (DIS) 
(Fig.~\ref{fig-feynman}c). 
Any leptoquark signal would have to be 
identified as a deviation from this SM prediction.
At high $M_{ljs}$, the SM prediction falls rapidly due to 
the dependence of the DIS cross sections on $Q^2$,
the virtuality of the exchanged boson, and to the
sharply falling valence-quark density at large Bjorken $x$. 
This  makes the data especially sensitive to virtual leptoquark exchange
and LQ-DIS interference.

Leptoquarks have been searched for previously 
in $ep$ collisions~\cite{epj:c16:253,pr:d63:052002,pr:d68:052004,Collaboration:2011qaa} 
and in $e^+e^-$~\cite{Acciarri:2000uh,Abbiendi:1998ea}, $p\bar{p}$~\cite{:2009gf,Acosta:2005ge} and $pp$~\cite{Aad:2011ch,Khachatryan:2010mp,Chatrchyan:2011ar} collisions.
Using $ep$ collisions, the Yukawa coupling, $\lambda$, of possible LQ
states to electron and electron neutrino
is probed. In $p\bar{p}$ and $pp$ collisions, the LQ production proceeds via the strong interaction and is independent of $\lambda$.
Thus the experimental approaches are complementary and $ep$ collisions
provide a unique testing ground.  
For this paper, the predictions for LQ cross sections were determined in leading order (LO) using the CTEQ5D parton density functions~\cite{epj:c12:375} (PDFs) using the Buchm\"uller-R\"uckl-Wyler model~\cite{pl:b191:442}. 
This model assumes that some of the leptoquarks are doublets or triplets
with degenerate masses. This differs from the assumptions made for
production via the strong interaction where in general a singlet state is assumed. The LHC experiments so far provided only limits for scalar LQs~\cite{Aad:2011ch,Khachatryan:2010mp,Chatrchyan:2011ar}. 

In the analysis presented here,
no evidence for any leptoquark signal was found. 
Therefore limits on $\lambda$ were derived as a function of $M_\mathrm{LQ}$ 
for the different leptoquark states described by the 
Buchm\"uller-R\"uckl-Wyler model. 

The analysis is based on the data collected by the ZEUS experiment 
in the period 2003--2007, corresponding to an integrated luminosity of \unit[366]{$\pbi$}. During this period, HERA was operated with a polarised lepton beam.  
The four data subsamples with different polarisation and lepton beam charge 
are summarised in Table~\ref{tab:lumi}. The experimental set-up described in Section~\ref{sec-exp} is that corresponding to these subsamples. 
The final limits set also included data collected in 1994--2000, giving a total integrated luminosity of \unit[498]{$\pbi$}. 
Thus all data from ZEUS were included and
the results presented here supersede 
those published previously~\cite{epj:c16:253,pr:d63:052002,pr:d68:052004}.

\section{Experimental set-up}
\label{sec-exp}

\Zdetdesc

\Zctdmvddesc{\ZcoosysfnBEeta}

\Zcaldesc



\Zlumidesc

The lepton beam in HERA became naturally transversely polarised
through the Sokolov-Ternov effect~\cite{Sokolov:1963zn,Baier:1969hw}. 
The characteristic build-up time for the HERA accelerator was
approximately 40~minutes.  Spin rotators on either side of the ZEUS
detector changed the  transverse polarisation of the beam into
longitudinal polarisation and back again. The lepton beam
polarisation was measured using  two independent polarimeters, the
transverse polarimeter (TPOL)~\cite{nim:a329:79} and the longitudinal
polarimeter (LPOL)~\cite{nim:a479:334}.  Both devices exploited the
spin-dependent cross section for Compton scattering of circularly
polarised photons off electrons to measure the beam polarisation.  The
luminosity and polarisation measurements were made over time intervals that
were much shorter than the polarisation build-up time. The polarisation values measured in different data taking periods are shown in Table~\ref{tab:lumi}. The fractional systematic uncertainty on the measured polarisation was 4.2\% for TPOL and 3.6\% for LPOL.

\section{Monte Carlo simulation}
\label{mc_simulation}

Monte Carlo techniques were used to determine the SM DIS background and the $M_{ljs}$ resolution of a possible signal. 

Standard Model NC and CC DIS events were simulated using the 
{\sc Heracles} 4.6.6~\cite{cpc:69:155,*spi:www:heracles} program 
with the {\sc Djangoh} 1.6~\cite{spi:www:djangoh11}
interfaces to the hadronisation programs and using CTEQ5D~\cite{epj:c12:375} PDFs. 
Radiative corrections for initial- and final-state 
electroweak radiation, vertex and propagator corrections,
and two-boson exchange were included.
The colour-dipole model of {\sc
  Ariadne} 4.12~\cite{cpc:71:15} was used to simulate
$\mathcal{O}(\alpha_{S})$ plus leading-logarithmic corrections to the
result of the quark-parton model.  {\sc Ariadne} uses the Lund string model 
of {\sc Jetset} 7.4.1~\cite{cpc:39:347,*cpc:43:367,*cpc:82:74} for the 
hadronisation.

The production and decay of resonances 
were simulated using PYTHIA 6.1~\cite{cpc:135:238},
which takes into account the finite width of
the resonant state, but includes only the $s$-channel diagrams. 
It also takes into account initial-
and final-state QCD radiation from the quark and the effect of LQ 
hadronisation 
before decay~\cite{Friberg:1997nn} as well as the initial-state
QED radiation from the electron. Such simulated samples of LQ events were used to study the bias and the resolution for the reconstructed LQ mass. The prediction for a LQ contribution to the DIS samples was evaluated by reweighting the DIS samples according to the LQ production processes (Section~\ref{sec_limits}). 

The ZEUS detector response was simulated using a program based on {\sc
  Geant} 3.21~\cite{tech:cern-dd-ee-84-1}.  The generated events were
passed through the detector simulation, subjected to the same trigger
requirements as the data and processed by the same reconstruction
programs.

\section{Leptoquark signal expectation}
\label{signal_exp}

The Buchm\"uller-R\"uckl-Wyler (BRW) model~\cite{pl:b191:442}
was used to calculate the cross sections in LO using the CTEQ5D PDFs for the LQ states listed
in Table~\ref{tab:brwlq}. The next-to-leading-order (NLO) QCD corrections,
the so-called $K$-factors~\cite{zfp:c74:611,zfp:c75:453}, 
available for scalar LQs were not considered, because equivalent calculations are not available for vector LQs. All limits presented in this paper are for LQ production in this model.

As the final states for LQ production are identical to
states produced in DIS, the DIS Monte Carlo samples were reweighted
accordingly to produce predictions for the respective $M_{ljs}$ spectra.
The BRW model predicts a dependence of the cross sections on the
beam polarisation. Therefore predictions were computed taking into account
the average polarisation of the respective data samples.

The BRW model includes both $u$ and $s$~channel and interferences with
DIS.
For $M_\mathrm{LQ} > \sqrt{s}$, the full LQ cross sections were used. 
For the virtual exchange, the cross section has a $\lambda^4$
dependence. 
The interference terms provide a contribution with a $\lambda^2$ dependence.
For $M_\mathrm{LQ} < \sqrt{s}$, some simplifications were introduced.
While the interference terms were calculated as for large $M_\mathrm{LQ}$,
the $u$-channel contribution, expected to be small, was neglected and  
the narrow-width approximation (NWA) was used for the resonant 
$s$-channel LQ production. 

The Born-level cross section for the resonant ($s$-channel) LQ production 
in the NWA is:

\begin{equation}
\sigma^{\mathrm{NWA}}=(J+1)\frac{\pi}{4s}\lambda^{2}\,q(x_{0}, M_\mathrm{LQ}^2) (1 \pm P_e),
\notag
\end{equation}

where $q(x_{0}, M_\mathrm{LQ}^2)$ is the initial-state quark (or antiquark) 
parton-density function in the proton for $x_{0}=M_\mathrm{LQ}^{2}/s$, 
$J$ is the spin of the LQ and the term $1 \pm P_e$ accounts for the dependence on the beam polarisation. In $e^- p$ ($e^+ p$) scattering, the polarisation dependence is given by $1+P_e$ ($1-P_e$) for LQs coupling to right-handed fermions and $1-P_e$ ($1+P_e$) for LQs coupling to left-handed fermions.  

The expected width of a LQ state with a mass between 100 and \unit[300]{\Gev} and a Yukawa coupling $\lambda=0.1$ ranges from 0.01 to \unit[0.2]{\Gev}. This justifies the use of the NWA for the $s$-channel contribution. To simulate a LQ signal in the MC, the $s$-channel term was added to the interference terms and the DIS MC events were reweighted. To reduce statistical fluctuations, the prediction from the NWA was smeared with a Gaussian narrower than the experimental resolution before adding the interference terms.

The effect of QED initial-state radiation, which decreases the production cross section, was taken into account for both resonant and non-resonant LQ production. The effect is larger for resonant LQ production and ranges up to $25\%$ for $M_\mathrm{LQ}$ close to $\sqrt{s}$.

The polarisation dependence is expected to be different for 
LQ production and DIS. The separation of the data according to
polarisation therefore provides a handle to identify a possible 
LQ signal.

Another possibility to isolate a leptoquark signal is the angular dependence
of the process.
The variable $\theta^*$, the lepton scattering 
angle in the lepton-jets centre-of-mass frame,
can be used to improve the signal-to-background ratio, 
especially for resonance production. 
The decay of a scalar resonance, for example,
will result in a flat distribution in $\cos \theta^*$, while NC DIS events
show approximately a $1/(1-\cos\theta^*)^2$ distribution.

\section{Signal search}
\label{sec:resonance_search}

Events from a hypothetical
resonance decaying
into $eq$ ($\nu$$q$) have a
topology identical to DIS $\NC$ ($\CC$) events.
Hence the final state from a high-mass resonance is expected to have at least one jet and either an identified final-state electron
or large missing transverse momentum.
The lepton-jet invariant mass was calculated as
\begin{equation}
M_{ljs} = \sqrt{E_{ljs}^2 - \vec{p}_{ljs}^{~2}},
\label{eq-Mljs}
\end{equation}
where $E_{ljs}$ is the sum of the energies of the outgoing lepton and the selected jets and $\vec{p}_{ljs}$ is the vector sum of the lepton and jets momenta. The modulus of the transverse momentum, $p_T$, and the net transverse energy, $E_T$, are defined as

\begin{alignat}{2}
p_T^2 & = & p_X^2 + p_Y^2 = & \left( \sum\limits_{i} E_i \sin \theta_i \cos
\phi_i \right)^2+ \left( \sum\limits_{i} E_i \sin \theta_i \sin \phi_i
\right)^2,
\label{eq-PT2}\\
E_T & = & \sum\limits_{i} E_i \sin \theta_i, \nonumber
\end{alignat}
where the sum runs over all calorimeter energy deposits, $E_i$.
The polar and azimuthal angles, $\theta_i$ and $\phi_i$,
of the calorimeter energy deposits were measured relative to the reconstructed event vertex. 
The quantity $E-p_Z$, also used in the event selection, is defined as

\begin{equation}
E-p_Z \equiv \sum\limits_{i} (E-p_Z)_{i} = \sum\limits_{i} ( E_i - E_i \cos
\theta_{i} ).
\label{eq-Delta}
\end{equation}

The hadronic polar angle was calculated as~\cite{Huttmann:2009zz}
\begin{equation}
\cos\gamma_h = \frac{P_{T,\mathrm{had}}^2-(E-P_Z)_{\mathrm{had}}}{P_{T,\mathrm{had}}^2+(E-P_Z)_{\mathrm{had}}}, 
\label{eq-gamma_h}
\end{equation}
where $P_{T,\mathrm{had}}^2$ and $(E-P_Z)_{\mathrm{had}}$ are calculated as in Equations~(\ref{eq-PT2}) and (\ref{eq-Delta}), but with the sum running only over the calorimeter energy deposits belonging to the hadronic final state. In case of the CC topology, $P_{T,\mathrm{had}}^2$ and $(E-P_Z)_{\mathrm{had}}$ are equivalent to $p_T^2$ and $E-p_Z$ from Equations~(\ref{eq-PT2}) and~(\ref{eq-Delta}), respectively.

\subsection{Neutral current, \boldmath$ep \to eX$\unboldmath, topology}
\label{nc_measurement}

\subsubsection*{Event selection}
\label{sec:event_select_nc}

The double-angle (DA) method~\cite{proc:hera:1991:23,*proc:hera:1991:43}
was used to reconstruct the kinematic variables.
In this method, the polar angle of the scattered electron and the hadronic polar angle (see Eq. (\ref{eq-gamma_h})) were used 
to reconstruct the kinematic variables $x_\DA$, $y_\DA$, and $Q^2_\DA$. The inelasticity $y$ was also reconstructed as $y_e$, using the electron method~\cite{proc:hera:1991:23,*proc:hera:1991:43}.

Events with the topology $ep \to eX$, where
$X$ denotes one or more jets,
were selected using the following criteria:

\begin{itemize}

\item the $Z$ coordinate
of the reconstructed event 
vertex was required to be in the
range $|Z|<\unit[30]{\cm}$, consistent with an $ep$ collision;

\item an electron identified in the CAL~\cite{epj:c11:427} was required. 
If the electron was found
within the acceptance of the tracking detectors, 
a track matched to the energy deposit in the calorimeter
was required with the distance of closest approach 
between the track extrapolated to the calorimeter surface
and the energy cluster position to be less than \unit[10]{\cm} 
and the electron track momentum, $p_e^{\rm trk}$, to be larger than \unit[3]{\Gev}.
A matched track was not required
if the electron emerged at a polar angle
outside the acceptance of the tracking detector.
Instead the electron was required to have a transverse momentum greater than \unit[30]{\Gev}.
An isolation requirement was imposed such that
the energy not associated with the electron
in an $\eta-\phi$ cone of radius 0.8 centred on the electron
was less than \unit[5]{\Gev};

\item a fiducial-volume cut was applied to the electron to guarantee that the experimental
acceptance was well understood. It excluded the transition regions between the FCAL
and the BCAL. It also excluded the regions within \unit[1.5]{\cm} of the module gaps in
the BCAL;

\item  at least one hadronic jet with transverse momentum 
$p^{j}_{T}>\unit[15]{\Gev}$ and $|\eta| < 3$, obtained
using the 
$k_T$ cluster algorithm~\cite{Catani:1993hr} in its longitudinally invariant 
inclusive mode~\cite{pr:d48:3160}, was required.
The centroid of any jet at the FCAL face was required to be outside a box of
$40\times \unit[40]{\cm}^2$ centred on the proton beam~\cite{inderpal:phd:2011}, 
in order to ensure good energy containment and to reduce
the systematic uncertainties due to the proton remnant. 
Additional jets were required to have $p^{j}_{T}>\unit[10]{\Gev}$ and $|\eta| < 3$;

\item to restrict the phase space to the region most relevant to the LQ search, cuts on $Q_\DA^2 > \unit[2500]{\Gev}^2$ and $x_\DA > 0.1$ were applied. Higher values of $x$ and $Q^2$ correspond to higher LQ masses, where the signal-to-background ratio for leptoquark events is higher;

\item to avoid phase-space regions in which the DIS MC generator was not valid, the quantity $y_\DA (1-x_\DA)^2$ was required to be larger than 0.004;

\item to remove background from photoproduction events and beam-gas events overlaid on NC events, 
the requirements $38 < E-p_{Z} < \unit[65]{\Gev}$ and $y_{e} < 0.95$ were imposed. 
To remove cosmic-ray events and beam-related background events, 
the quantity $p_{T}/\sqrt{E_{T}}$ was required to be less than $4\sqrt{\gev}$
and the quantity $p_{T}/E_{T}$ was required to be less than 0.7. Elastic QED Compton events were rejected by considering the balance between the photon and the electron~\cite{Yongdok}.

\end{itemize}

The mass shifts and resolutions for resonant lepton-quark states 
were calculated from the LQ MC. The mass resolution,
determined from a Gaussian fit to the peak of the reconstructed mass
spectrum, fell from 5\% to 3\% as the resonant mass increased
from 150 to \unit[290]{\Gev}. Any mass shift was within 
0.5\% for LQ masses between
150 and \unit[290]{\Gev}.

\subsubsection*{Search results}

\begin{figure}[p]
\vfill
\begin{center}
\includegraphics[width=11cm]{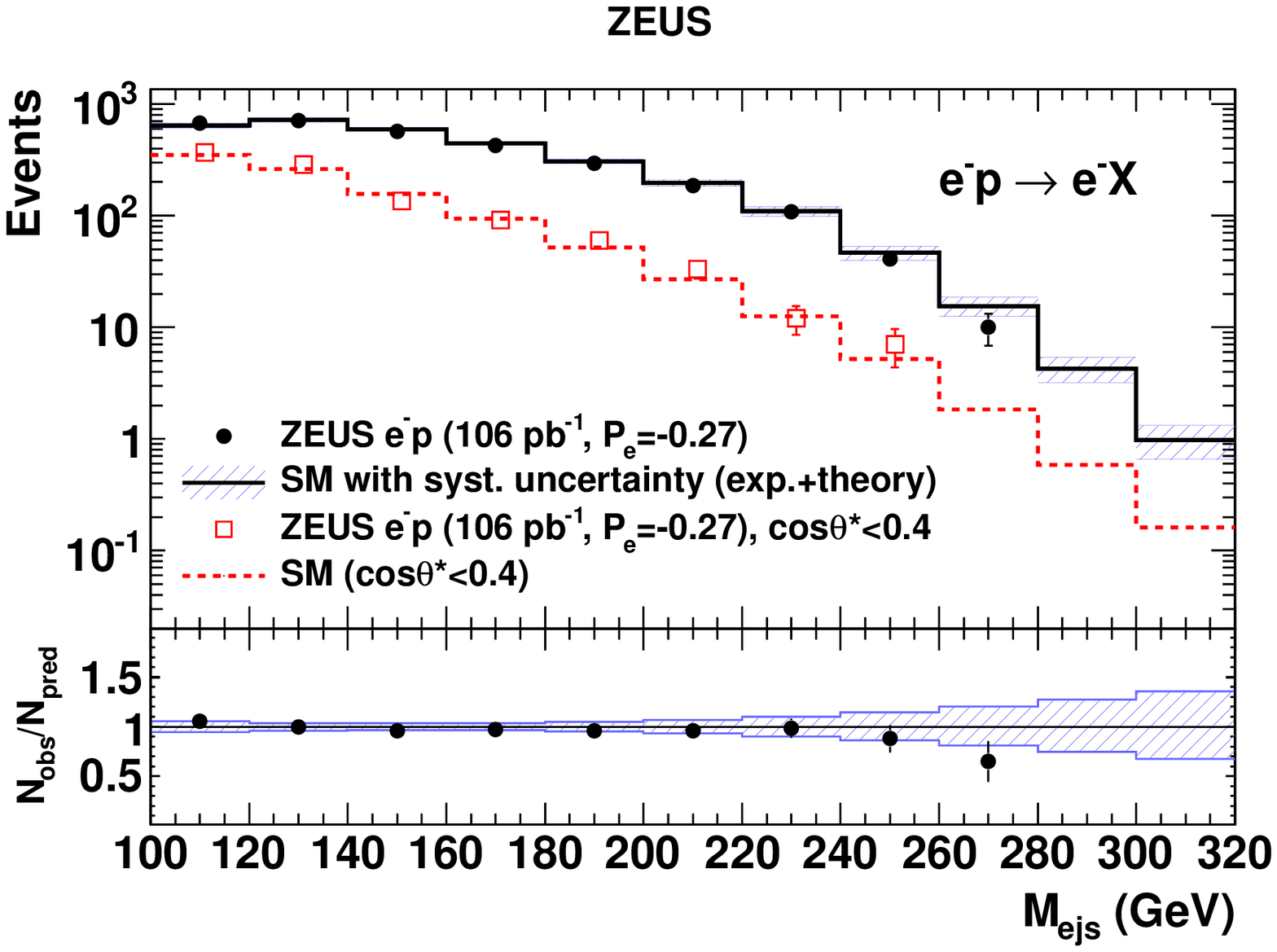}
\end{center}
\caption{
Comparison of 
the left-handed $e^-p$ sample (dots) and the NC SM 
expectation (solid histogram) for the 
reconstructed invariant mass, $M_{ejs}$, in
the $e^-p \to e^-X$ topology. 
The data (open squares) and the SM expectation (dashed
histogram) for $\cos\theta^*<$0.4 are also shown.
The shaded area shows the overall uncertainty of the SM MC expectation. 
The lower part of the plot shows the ratio between the data and the SM expectation without the $\cos{\theta^*}$ cut. }
\label{fig-Mass_NC_Ele_LH}
\vfill
\end{figure}

\begin{figure}[p]
\vfill
\begin{center}
\includegraphics[width=11cm]{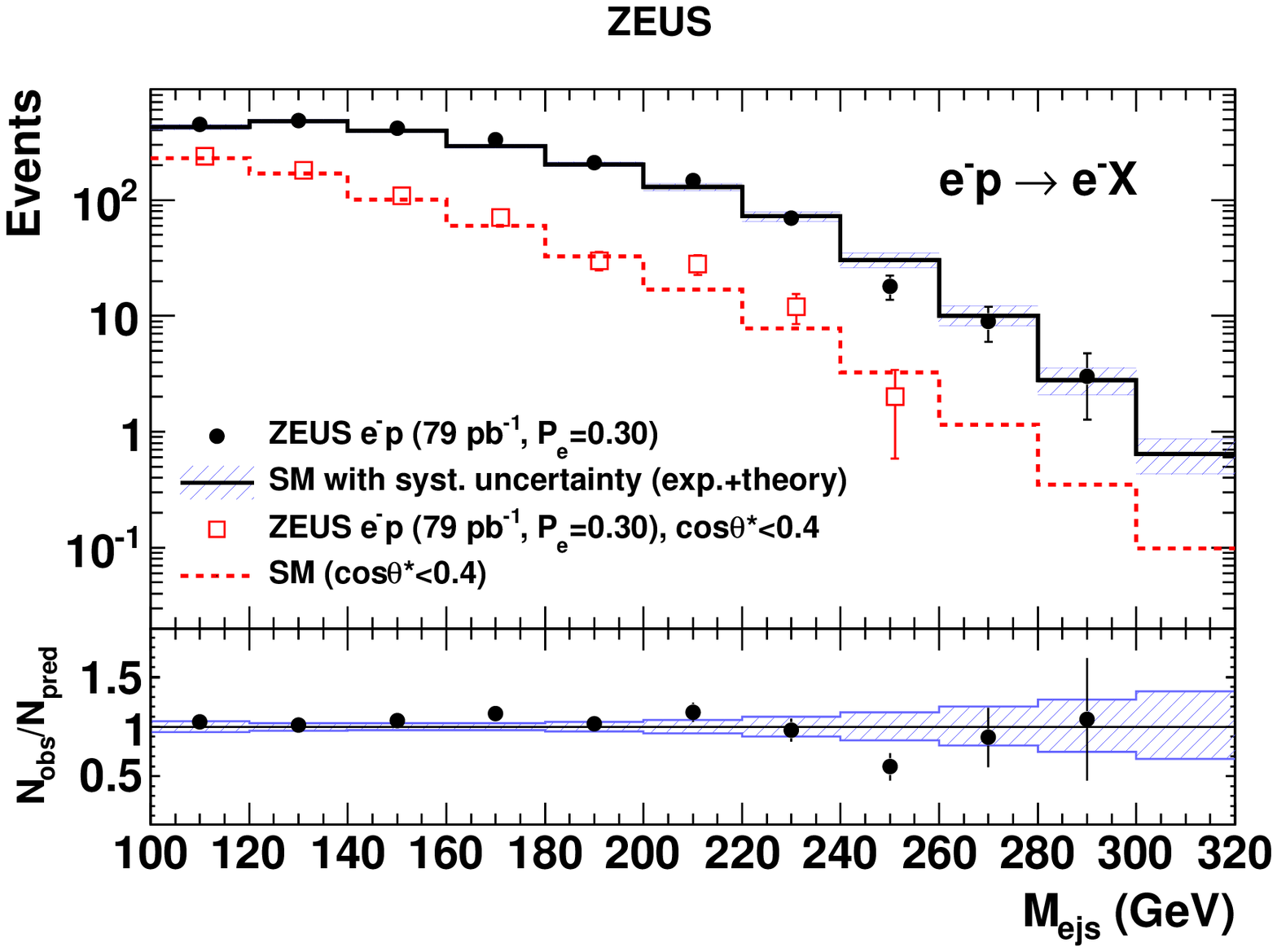}
\end{center}
\caption{
Comparison of 
the right-handed $e^-p$ sample (dots) and the NC SM 
expectation (solid histogram) for the 
reconstructed invariant mass, $M_{ejs}$, in
the $e^-p \to e^-X$ topology. 
Other details as in the caption to Fig.~\ref{fig-Mass_NC_Ele_LH}.}
\label{fig-Mass_NC_Ele_RH}
\vfill
\end{figure}

\begin{figure}[p]
\vfill
\begin{center}
\includegraphics[width=11cm]{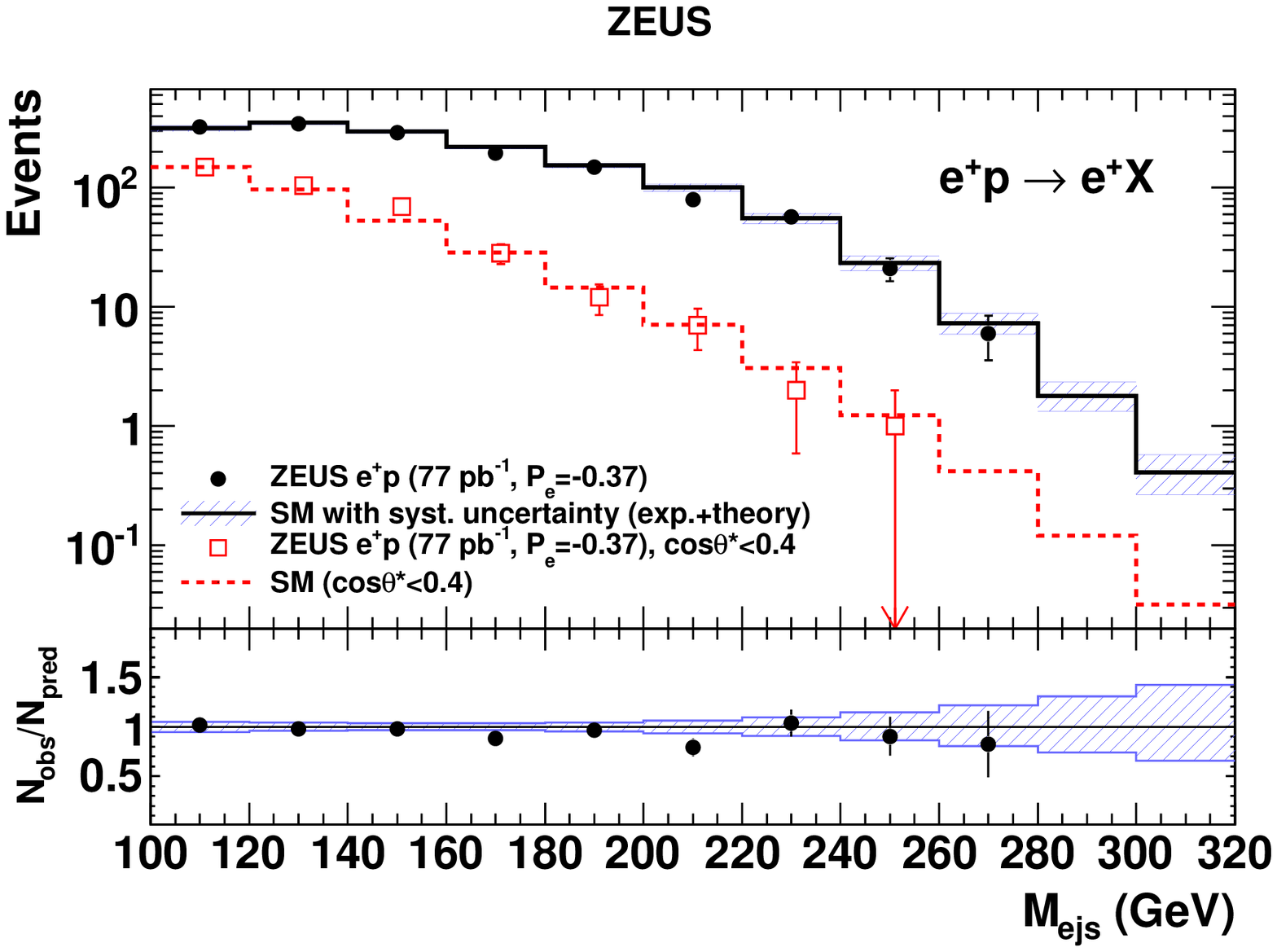}
\end{center}
\caption{
Comparison of 
the left-handed $e^+p$ sample (dots) and the NC SM 
expectation (solid histogram) for the 
reconstructed invariant mass, $M_{ejs}$, in
the $e^+p \to e^+X$ topology. 
Other details as in the caption to Fig.~\ref{fig-Mass_NC_Ele_LH}.}
\label{fig-Mass_NC_Posi_LH}
\vfill
\end{figure}

\begin{figure}[p]
\vfill
\begin{center}
\includegraphics[width=11cm]{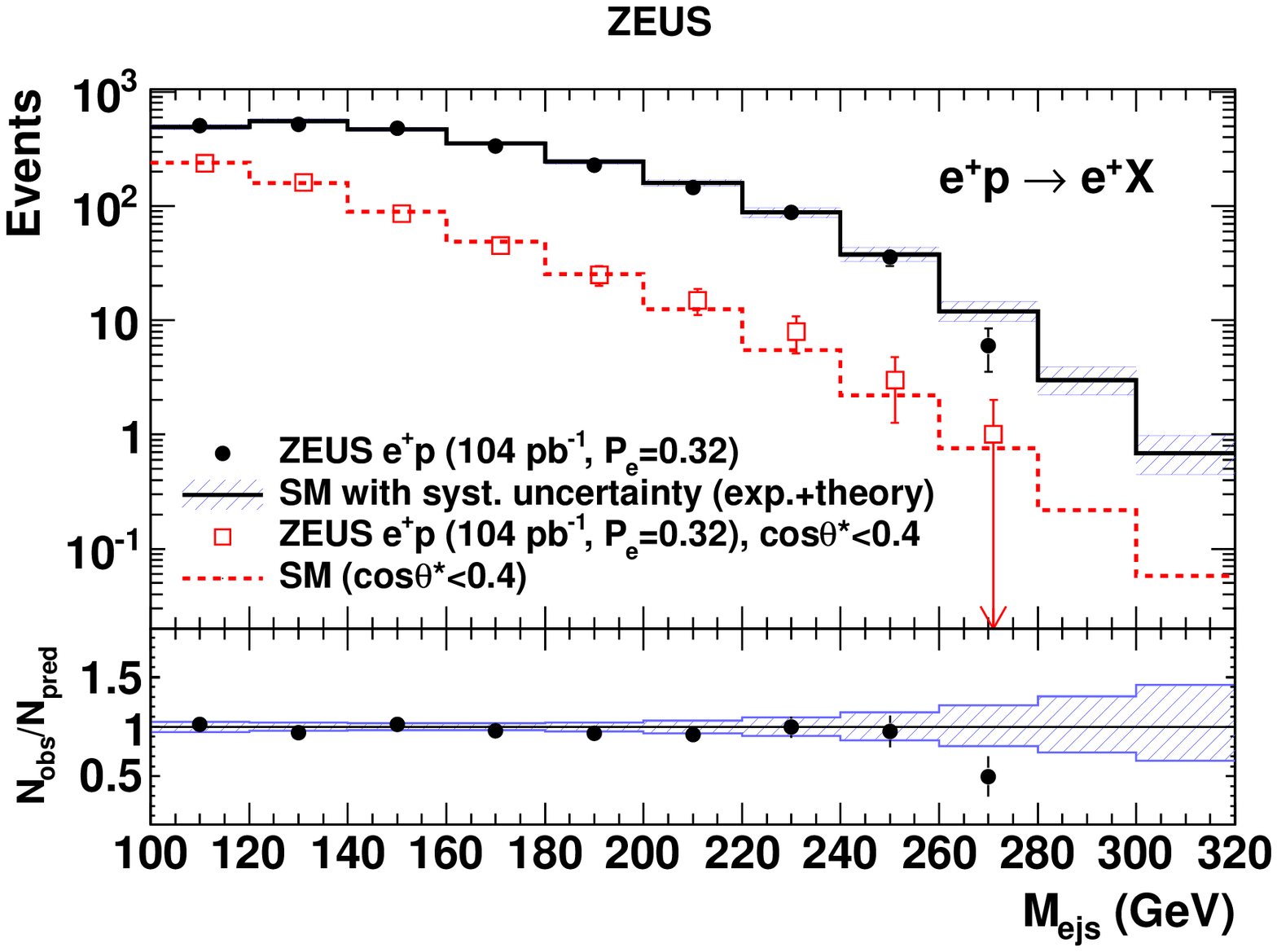}
\end{center}
\caption{
Comparison of 
the right-handed $e^+p$ sample (dots) and the NC SM 
expectation (solid histogram) for the 
reconstructed invariant mass, $M_{ejs}$, in
the $e^+p \to e^+X$ topology. 
Other details as in the caption to Fig.~\ref{fig-Mass_NC_Ele_LH}.}
\label{fig-Mass_NC_Posi_RH}
\vfill
\end{figure}

\begin{figure}[p]
\vfill
\begin{center}
\includegraphics[width=11cm]{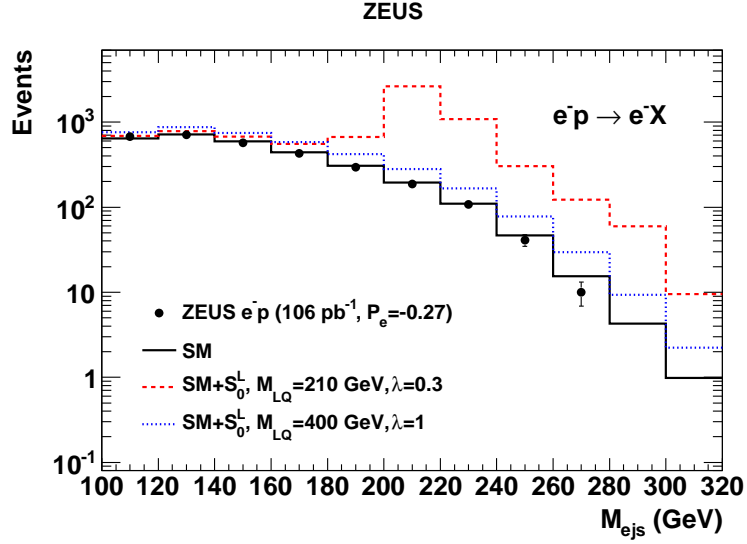}
\end{center}
\caption{
Comparison of the reconstructed invariant mass, $M_{ejs}$, distribution in the $e^-p \to e^-X$ topology for the left-handed $e^-p$ sample (dots) to the NC SM 
expectation (solid histogram) and to the predictions of the model including a $S_0^L$ LQ state with a mass of 210~GeV and a coupling $\lambda$ of 0.3 (dashed histogram) as well as a mass of 400~GeV and a coupling $\lambda$ of 1 (dotted histogram).}
\label{fig-Mass_LQSignal_2}
\vfill
\end{figure}

\begin{figure}[p]
\vfill
\begin{center}
\includegraphics[width=11cm]{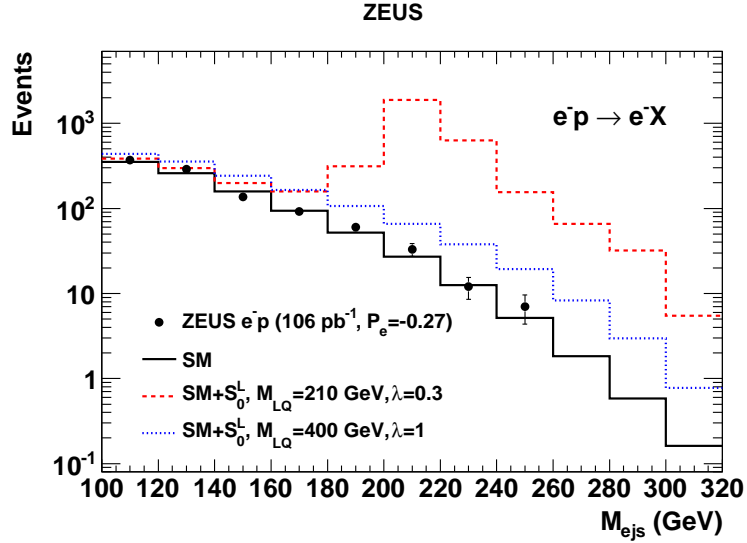}
\end{center}
\caption{
As Fig.~\ref{fig-Mass_LQSignal_2}, but the cut on $\cos\theta^*<$0.4 was applied.}
\label{fig-Mass_LQSignal_cost_2}
\vfill
\end{figure}

After the above selection, 9\,369 events were found in the data from 2003--2007, 
compared to 9\,465 $\pm$ 494 
expected from the $\NC$ MC 
and the evaluation of its systematic uncertainties (see below). 
The measured distributions of the $M_{ejs}$ spectra for $e^-p$ ($e^+p$) data with a left-handed and a right-handed lepton beam are shown in Figs.~\ref{fig-Mass_NC_Ele_LH} and \ref{fig-Mass_NC_Ele_RH} (Figs.~\ref{fig-Mass_NC_Posi_LH} and \ref{fig-Mass_NC_Posi_RH}), respectively. A cut on $\cos\theta^*<0.4$, introduced to suppress further the SM background~\cite{epj:c16:253,pr:d63:052002}, was also applied. The upper parts of the plots show 
the spectra with and without 
the $\cos\theta^*$ cut, 
while the lower parts show the ratio
of the observed spectrum to SM expectations with no cut applied on $\cos\theta^*$. Good agreement is seen between the
data and the SM NC prediction. 
Figures~\ref{fig-Mass_LQSignal_2} and~\ref{fig-Mass_LQSignal_cost_2} show the $e^-p$ data for the left-handed electron beam together with the predictions for a $S_0^L$ LQ state with a mass of \unit[210]{\Gev} and a coupling $\lambda$ of 0.3 as well as a mass of \unit[400]{\Gev} and a coupling $\lambda$ of 1 without and with a $\cos{\theta^*}$ cut, respectively.

\subsubsection*{Systematic uncertainties}
\label{sec:nc_systematic}

The uncertainty
on the expected number of events from SM NC DIS processes 
was investigated. Relevant were~\cite{Huttmann:2009zz}:

\begin{itemize}
\item the uncertainty on the calorimeter energy scale, 1\% for electrons and 2\% for hadrons. This led to an uncertainty of 1\% (6\%) in the NC expectation for $M_{ejs}$=150 \unit[(220)]{\Gev}; 

\item the uncertainty on the parton densities as
estimated by Botje~\cite{epj:c14:285}, which gave an uncertainty of 3\% (5\%) in the NC expectation 
for $M_{ejs}$=150 \unit[(220)]{\Gev};

\item the uncertainty on the luminosity determination of 1.8\% for electron and 2.2\% for positron beams\footnote{For a fraction of the positron data, the uncertainty was 3.5\%, while for most of the positron data the uncertainty was 1.8\%. This lead to a total uncertainty of 2.2\%.}, which is directly reflected in the result.

\end{itemize}

The overall systematic uncertainties 
on the background expectations were obtained by adding all relevant contributions in quadrature. They are 
shown as hatched bands in Figs.~\ref{fig-Mass_NC_Ele_LH}, \ref{fig-Mass_NC_Ele_RH}, \ref{fig-Mass_NC_Posi_LH} and \ref{fig-Mass_NC_Posi_RH}. For a given mass, the systematic uncertainty for the LQ signal is assumed to be the same as for the SM background.

\subsection{Charged current, \boldmath$ep \to \nu X$\unboldmath, topology}
\label{cc_measurement}

\subsubsection*{Event selection}
The events with the topology $ep \to \nu X$, where $X$ denotes one or more jets, are classified according to $\gamma_0$, the hadronic polar angle (see Eq.~(\ref{eq-gamma_h})) assuming a nominal vertex position of $Z = 0$. 
Events for which the hadronic system is not contained in the CTD acceptance ($\gamma_0 \le 0.4~\rad$) are called low-$\gamma_0$ events. The hadronic systems of high-$\gamma_0$ events with $\gamma_0 >0.4~\rad$ are inside the CTD acceptance, so that cuts based on tracking information can be applied. 
The kinematic variables were reconstructed using the Jacquet-Blondel
 method \cite{proc:epfacility:1979:391}. 

The events were selected using the following criteria: 

\begin{itemize}

\item the $Z$ coordinate
of the reconstructed event 
vertex was required to be in the
range $|Z|<\unit[30]{\cm}$, consistent with an $ep$ collision;

\item to restrict the phase space to the region most relevant to the LQ search, a cut on $Q_\JB^2 > \unit[700]{\Gev}^2$ was applied. Since the resolution on $Q^2_\JB$ was poor at high $y$, a cut on $y_\JB < 0.9$ was added; 

\item a missing transverse momentum $p_{T}>\unit[22]{\Gev}$ was required and,
  to suppress beam-gas events, the missing  transverse momentum excluding the
  calorimeter cells adjacent to the forward beam hole was
  required to exceed \unit[20]{\Gev}. Compared to the ZEUS CC DIS analyses~\cite{Chekanov:2006da,Chekanov:2008aa,Collaboration:2010xc}, these cuts are more stringent because CC events with low $p_T$ lead to low invariant masses of the LQs and are therefore not of interest in this analysis;

\item in the high-$\gamma_0$ region, cuts based on the number and quality of tracks were applied. Tracks with a transverse momentum above \unit[0.2]{\Gev} were selected. They were required to start from the MVD or the innermost superlayer of the CTD, and had to reach at least the third superlayer of the CTD. If in addition they pointed to the primary vertex, they were considered as "good tracks". At least one good track was required in the event and the ratio of the total number of tracks to the number of good tracks had to fulfill the requirement $\frac{N_\mathrm{trk} - 20}{N_\mathrm{gtrk}} < 4$, where $N_\mathrm{trk}$ is the total number of tracks and $N_\mathrm{gtrk}$ is the number of good tracks. This cut removed beam-gas events which are characterised by a high number of poor-quality tracks;

\item at least one hadronic jet with transverse momentum 
$p^{j}_{T}>\unit[10]{\Gev}$ and $|\eta| < 3$, obtained
using the 
$k_T$ cluster algorithm~\cite{Catani:1993hr} in its longitudinally invariant  
inclusive mode~\cite{pr:d48:3160}, was required.
The centroids of all jets at the FCAL face were required to be outside a box of
$40\times \unit[40]{\cm}^2$ centred on the proton beam~\cite{inderpal:phd:2011}, 
in order to ensure good energy containment and to reduce
the systematic uncertainties due to the proton remnant.

\item to reject photoproduction and di-lepton background, for events with $p_T<\unit[30]{\Gev}$ a dedicated cut based on the energy distribution in the detector relative to the total transverse momentum was used. The transverse momentum sum for the calorimeter cells with a positive contribution to the total transverse momentum (parallel component $V_{P}$) and the corresponding sum for cells giving a negative contribution to the total transverse momentum (antiparallel component $V_{AP}$) had to satisfy the condition $V_{AP}/V_{P}$ $<0.35$~\cite{Collaboration:2010xc}. 
This requirement demanded an azimuthally collimated energy flow.  In addition, for events with at least one good track, the azimuthal-angle difference, $\Delta\phi$, between the missing transverse  momentum measured by the tracks\footnote{The missing transverse momentum measured by the tracks is calculated using all the good tracks.} and that measured by the calorimeter was required to be less than $45^\circ$ for events with $p_T<\unit[30]{\Gev}$ and less than $60^\circ$ otherwise. This cut rejects events caused by cosmic rays or muons in the beam halo as well as beam-gas events;

\item NC events were removed by discarding events containing 
electron candidates with an energy greater than \unit[4]{\Gev}~\cite{Chekanov:2008aa}; 

\item requirements on energy fractions in the calorimeter cells plus muon-finding algorithms based on tracking, calorimeter and muon-chamber information were used to reject events caused by cosmic rays or muons in the beam halo. Furthermore, the deposition times of the energy clusters in the calorimeter were checked to be consistent with the bunch-crossing time to reject events due to interactions between the beams and residual gas in the beam pipe or upstream accelerator components. In addition, topological cuts on the transverse and longitudinal shower shape were imposed to reject beam-halo muon events that produced a shower inside the FCAL. Cuts on the calorimeter cell with the highest transverse energy were applied to reject sparks faking a CC event~\cite{Huttmann:2009zz}.

\end{itemize}

The neutrino energy and angle were calculated 
by assuming that missing $p_T$ and missing $E-P_Z$
were carried away by a single neutrino and used to calculate 
the invariant mass of the $\nu$-jets system,~$M_{\nu js}$, according to Eq.~(\ref{eq-Mljs}).

The shift and resolution of the invariant mass
were studied by using the LQ MC events and
fitting the mass peak with a Gaussian function.
The resulting mass shift was within 
0.5\% for LQ masses between
150 and \unit[290]{\Gev}, with the resolution varying from 8\% to 6\%, respectively.

\subsubsection*{Search results}

\begin{figure}[p]
\vfill
\begin{center}
\includegraphics[width=11cm]{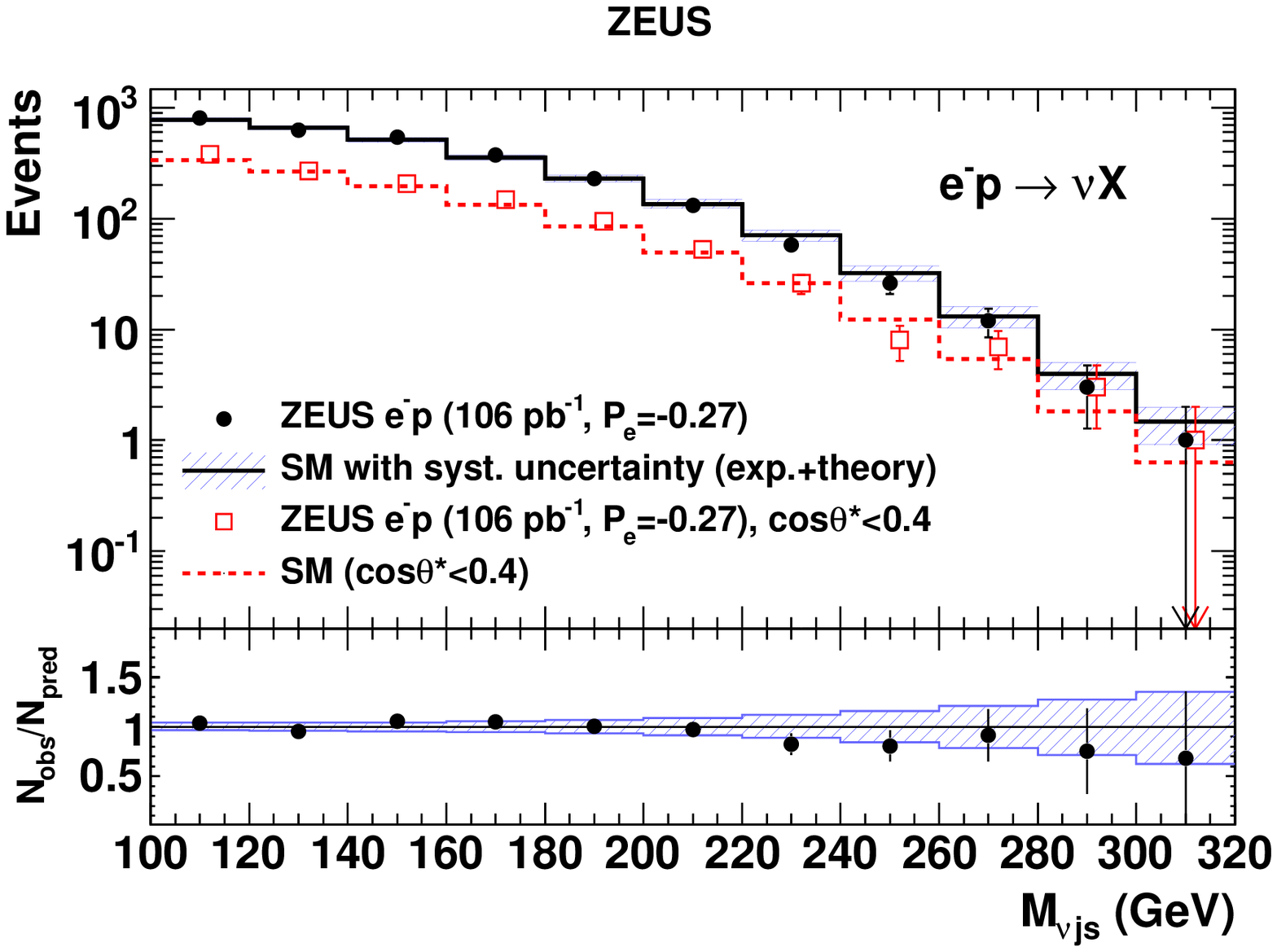}
\end{center}
\caption{
Comparison of 
the left-handed $e^-p$ sample (dots) and the CC SM 
expectation (solid histogram) for the 
reconstructed invariant mass, $M_{\nu js}$, in
the $e^-p \to \nu X$ topology. 
The data (open squares) and the SM expectation (dashed
histogram) for $\cos\theta^*<$0.4 are also shown.
The shaded area shows the overall uncertainty of the SM MC expectation. 
The lower part of the plot shows the ratio between the data and the SM expectation without the $\cos{\theta^*}$ cut. }
\label{fig-Mass_CC_Ele_LH}
\vfill
\end{figure}

\begin{figure}[p]
\vfill
\begin{center}
\includegraphics[width=11cm]{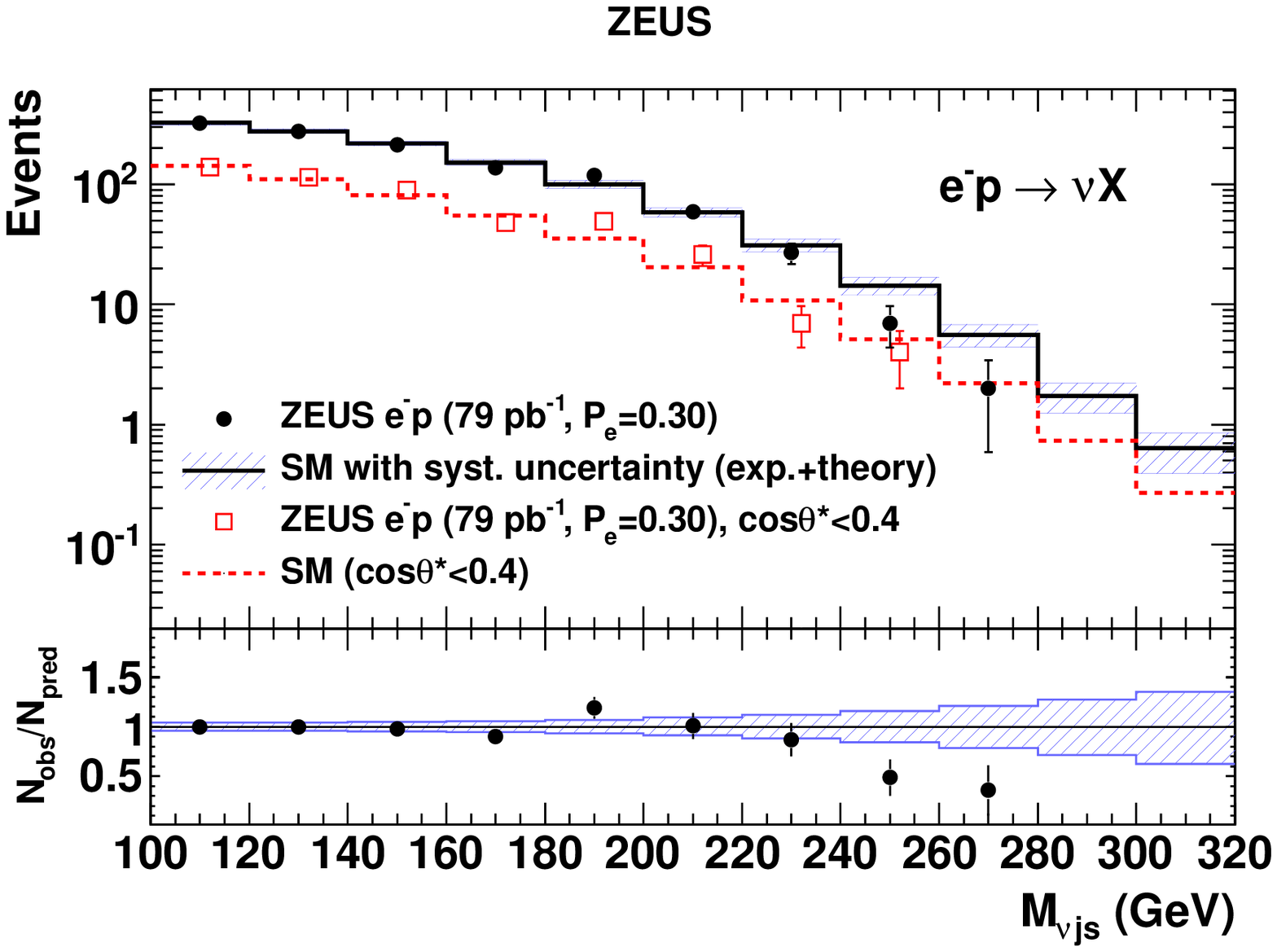}
\end{center}
\caption{
Comparison of 
the right-handed $e^-p$ sample (dots) and the CC SM 
expectation (solid histogram) for the 
reconstructed invariant mass, $M_{\nu js}$, in
the $e^-p \to \nu X$ topology. Other details as in the caption to Fig.~\ref{fig-Mass_CC_Ele_LH}.}
\label{fig-Mass_CC_Ele_RH}
\vfill
\end{figure}

\begin{figure}[p]
\vfill
\begin{center}
\includegraphics[width=11cm]{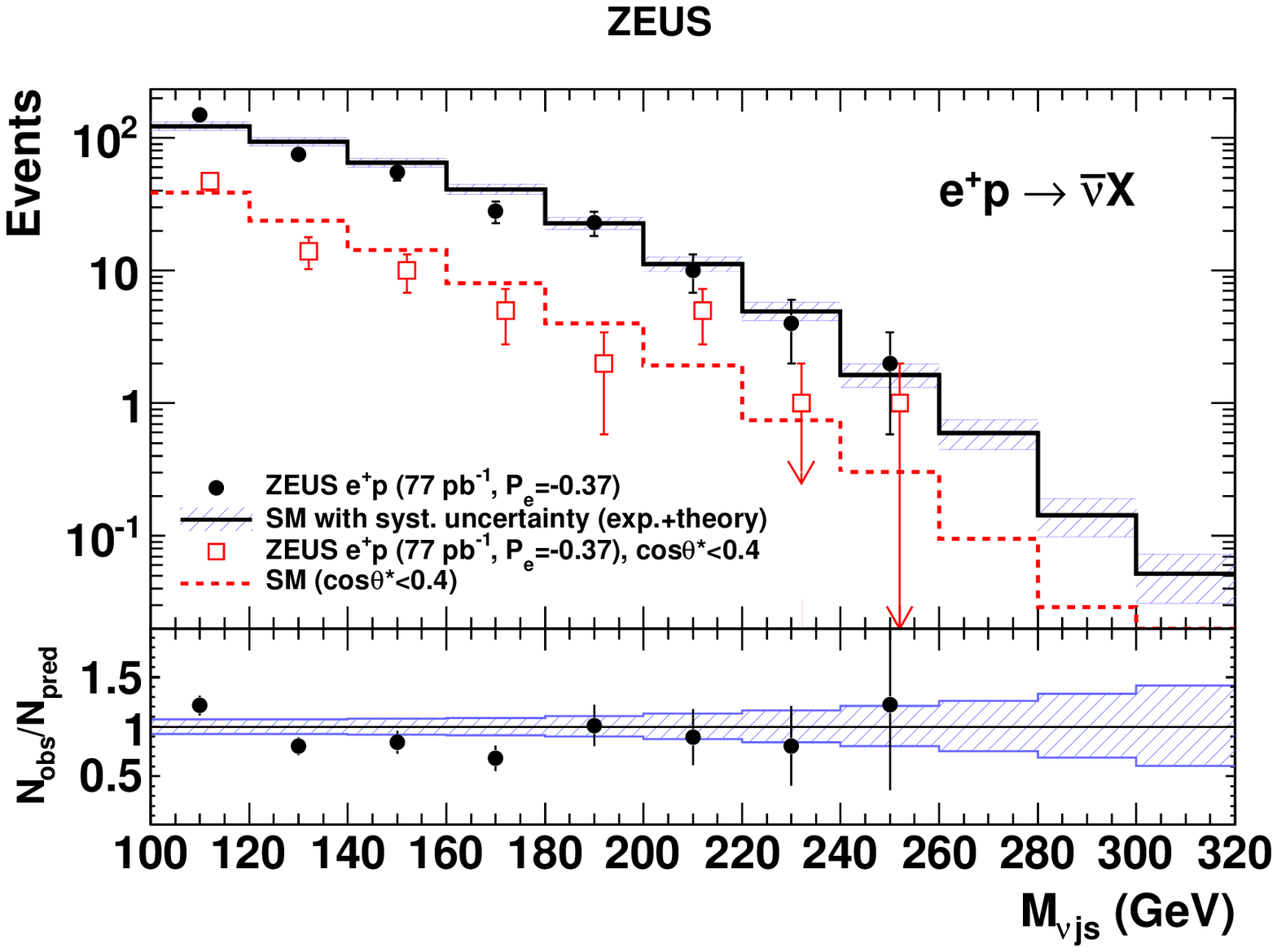}
\end{center}
\caption{
Comparison of 
the left-handed $e^+p$ sample (dots) and the CC SM 
expectation (solid histogram) for the 
reconstructed invariant mass, $M_{\nu js}$, in
the $e^+p \to \bar{\nu}X$ topology. 
Other details as in the caption to Fig.~\ref{fig-Mass_CC_Ele_LH}.}
\label{fig-Mass_CC_Posi_LH}
\vfill
\end{figure}

\begin{figure}[p]
\vfill
\begin{center}
\includegraphics[width=11cm]{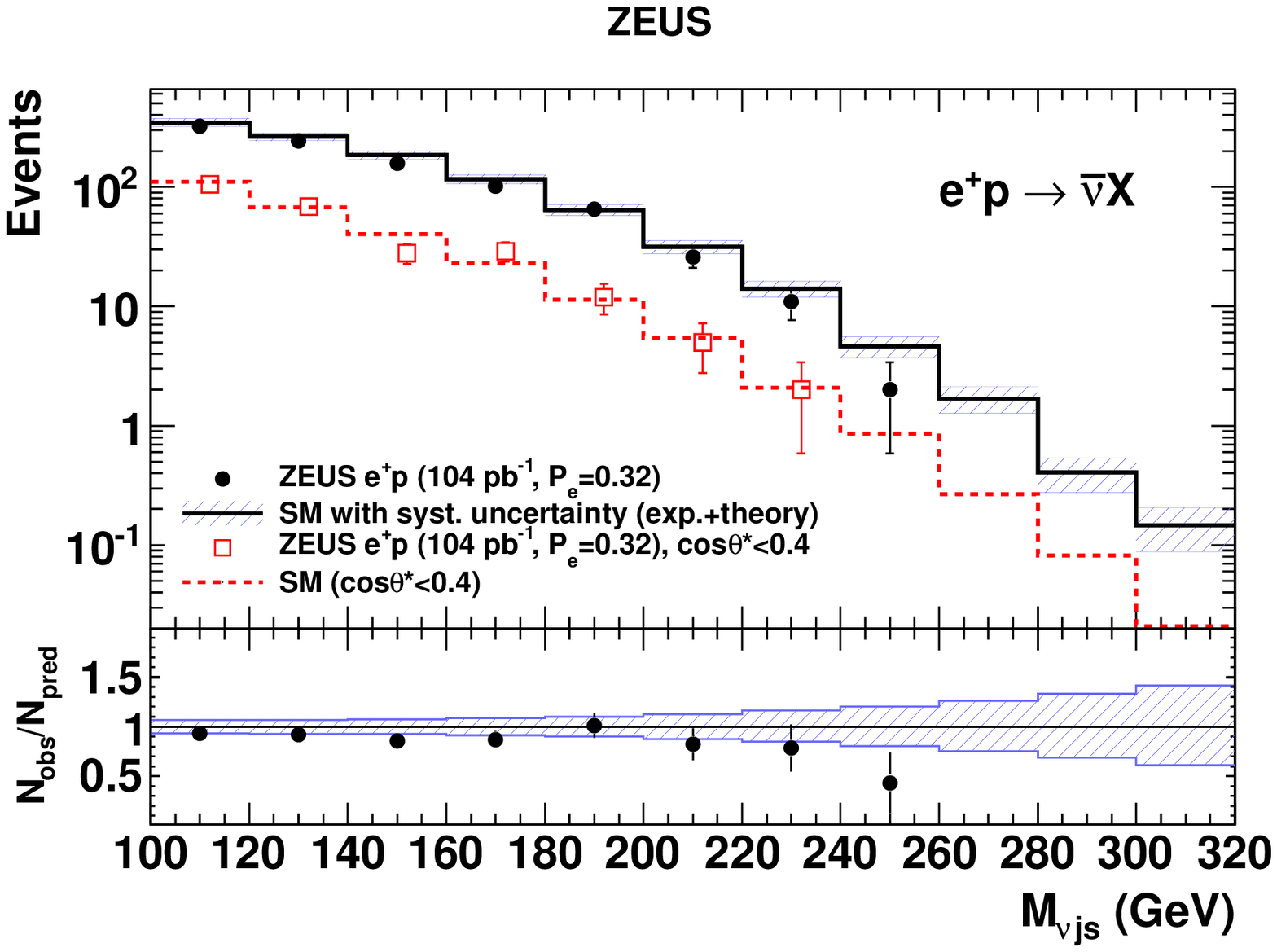}
\end{center}
\caption{
Comparison of 
the right-handed $e^+p$ sample (dots) and the CC SM 
expectation (solid histogram) for the 
reconstructed invariant mass, $M_{\nu js}$, in
the $e^+p \to \bar{\nu}X$ topology. 
Other details as in the caption to Fig.~\ref{fig-Mass_CC_Ele_LH}.}
\label{fig-Mass_CC_Posi_RH}
\vfill
\end{figure}

After the above selection, 8\,990 events were found in the data from 2003--2007, 
compared to 9\,068 $\pm$ 501 
expected from the $\CC$ MC 
and the evaluation of its systematic uncertainties (see below). 
The measured distributions of the $M_{\nu js}$ spectra for the left-handed and right-handed $e^-p$ ($e^+p$) data are shown in Figs.~\ref{fig-Mass_CC_Ele_LH} and \ref{fig-Mass_CC_Ele_RH} (Figs.~\ref{fig-Mass_CC_Posi_LH} and \ref{fig-Mass_CC_Posi_RH}). The upper parts of the plots show the spectra with and without the cut
$\cos\theta^*<0.4$, 
while the lower parts show the ratio
of the observed spectrum to SM expectations with no cut applied on $\cos\theta^*$. Good agreement is seen between the
data and the SM CC prediction. 

\subsubsection*{Systematic uncertainties}
\label{sec:cc_systematic}

The uncertainty on the predicted background from SM CC DIS processes
was investigated. The uncertainties found to be relevant~\cite{Huttmann:2009zz} are similar to those described in Section~\ref{sec:nc_systematic}
for the $ep \to eX$ case and 
arise from:
\begin{itemize}

\item the uncertainy on the hadronic energy scale of 2\%, which led to an uncertainty of 3\% (10\%) in the CC expectation for $M_{\nu js}$=150 \unit[(220)]{\Gev}; 

\item the uncertainty on the parton densities as estimated by Botje~\cite{epj:c14:285}, giving 3\% (4\%) and 7\% (9\%) uncertainties on the cross section for 
$e^-p$ and $e^+p$, respectively, 
for $M_{\nu js}$=150 \unit[(220)]{\Gev}. The correlations between $e^-p$ and $e^+p$ as well as NC and CC cross section uncertainties were taken into account; 

\item the uncertainty on the luminosity determination of 1.8\% for electron and 2.2\% for positron beams, which is directly reflected in the result;

\item the uncertainty on the measured polarisation of 4.2\%. To be conservative, the TPOL uncertainty, which is larger than the LPOL uncertainty, was used. This led to an uncertainty on the SM cross section of 0.9\% (2.4\%) for left-handed $e^-p$ ($e^+p$) data and 1.8\% (1.0\%) for right-handed $e^-p$ ($e^+p$) data, respectively.

\end{itemize}

The overall systematic uncertainties 
on the background expectations were obtained by adding all relevant contributions in quadrature. They are
shown as hatched bands in Figs.~\ref{fig-Mass_CC_Ele_LH}, \ref{fig-Mass_CC_Ele_RH}, \ref{fig-Mass_CC_Posi_LH} and \ref{fig-Mass_CC_Posi_RH}. For a given mass, the systematic uncertainty for the LQ signal is assumed to be the same as for the SM background.

\section{Limits on leptoquarks}
\label{sec_limits}

The expectation from a potential LQ signal was obtained by reweighting the DIS MC according to the cross sections predicted in the BRW model (see Section~\ref{signal_exp}). Each MC event is reweighted with the following weighting factor, WF:
\begin{equation}
\mathrm{WF}(x,y;P_e;M_\mathrm{LQ},\lambda)=\frac{{{\frac{d^2\sigma^{\mathrm{SM} +LQ}}{dxdy}}(x,y;P_e;M_\mathrm{LQ},\lambda)}}{{\frac{d^2\sigma^{\mathrm{SM}}}{dxdy}}(x,y;P_e)},
\notag
\end{equation}
where $x$ and $y$ are the true kinematic variables of the MC simulation, and $P_e$ is the average polarisation of the data sample given in Table~\ref{tab:lumi}. The effect of QED initial-state radiation was taken into account.

The limits were calculated including the results of the search presented here and the data recorded with the ZEUS detector in the years 1994--2000~\cite{pr:d68:052004}. 
They were set using a binned likelihood technique
 in the ($M_{ljs}$, $\cos\theta^*$) plane. The region 150 $<M_{ljs}<\unit[320]{\Gev}$ was used. 
The data were binned separately for each of the data sets listed in Table~\ref{tab:lumi}, thereby taking into account different beam charges and polarisation. 
For leptoquark states with $\nu q$ decays,
both the $eq \to eX$ and the $eq \to \nu X$ samples were used,
while for leptoquark states decaying only to $eq$,
only the $eq \to eX$ samples were used.

The upper limit on the coupling strength,
$\lambda_{\rm limit}$,
as a function of $M_\mathrm{LQ}$, was obtained by solving
\begin{equation}
\int_0^{\lambda_{\rm limit}^2}d\lambda^2
L(M_\mathrm{LQ},\lambda) = 
0.95 \int_0^{\infty}d\lambda^2
L(M_\mathrm{LQ}, \lambda),
\label{eq:lam_lim}
\end{equation}
where $L$ is the product of
the Poisson probabilities of the observed number of events in bin $i$, $N_i$, with the expectation, $\mu_i$, convoluted with Gaussian distributions for the main systematic uncertainties, yielding a modified expectation $\mu_i^\prime$: 
\begin{equation}
 L = \int_{-\infty}^{\infty} \prod_j d\delta_j 
   \frac{1}{\sqrt{2\pi}} e^{(-\delta_j^2/2)}
   \prod_i  e^{(-\mu'_i)} \frac{{\mu'_i}^{N_i}}{N_i!}.
\notag
\end{equation}
The index $j$ denotes the source of a systematic uncertainty and
$\delta_j$ corresponds to the variation of the
$j^{\rm th}$ systematic parameter in units of the nominal values
quoted in Sections \ref{sec:nc_systematic} and \ref{sec:cc_systematic}.
The index $i$ runs over the bins in $M_{ljs}$-$\cos\theta^*$ and
the relevant data sets. 
The modified expectation was calculated as  
\[ \mu'_i = \mu_i\prod_j (1+\sigma_{ij})^{\delta_j}, \]
where $\sigma_{ij}$ gives the fractional variation of $\mu_i$ under the
nominal shift in the $j^{\rm th}$ systematic parameter.
This \it ansatz \rm of $\mu'_i$ reduces to a linear dependence of
$\mu'_i$ on each $\delta_j$ when $\delta_j$ is small, while avoiding
the possibility of $\mu'_i$ becoming negative which would arise if
$\mu'_i$ was defined as a linear function of the $\delta_j$.

The coupling limits for 
the 14 BRW LQs listed in Table~\ref{tab:brwlq} were calculated
for masses up to \unit[1]{\Tev}. 
For large LQ masses, i.e. $M_\mathrm{LQ} \gg \sqrt{s}$, their exchange can be described as a contact interaction with an effective coupling proportional to $\lambda^2/M_\mathrm{LQ}^2$. Table~\ref{limit_table_1TeV} shows the limits on $\lambda$ for all BRW LQs at $M_\mathrm{LQ} = \unit[1]{\Tev}$. 

\begin{figure}[p]
\vfill
\begin{center}
\includegraphics[width=11cm]{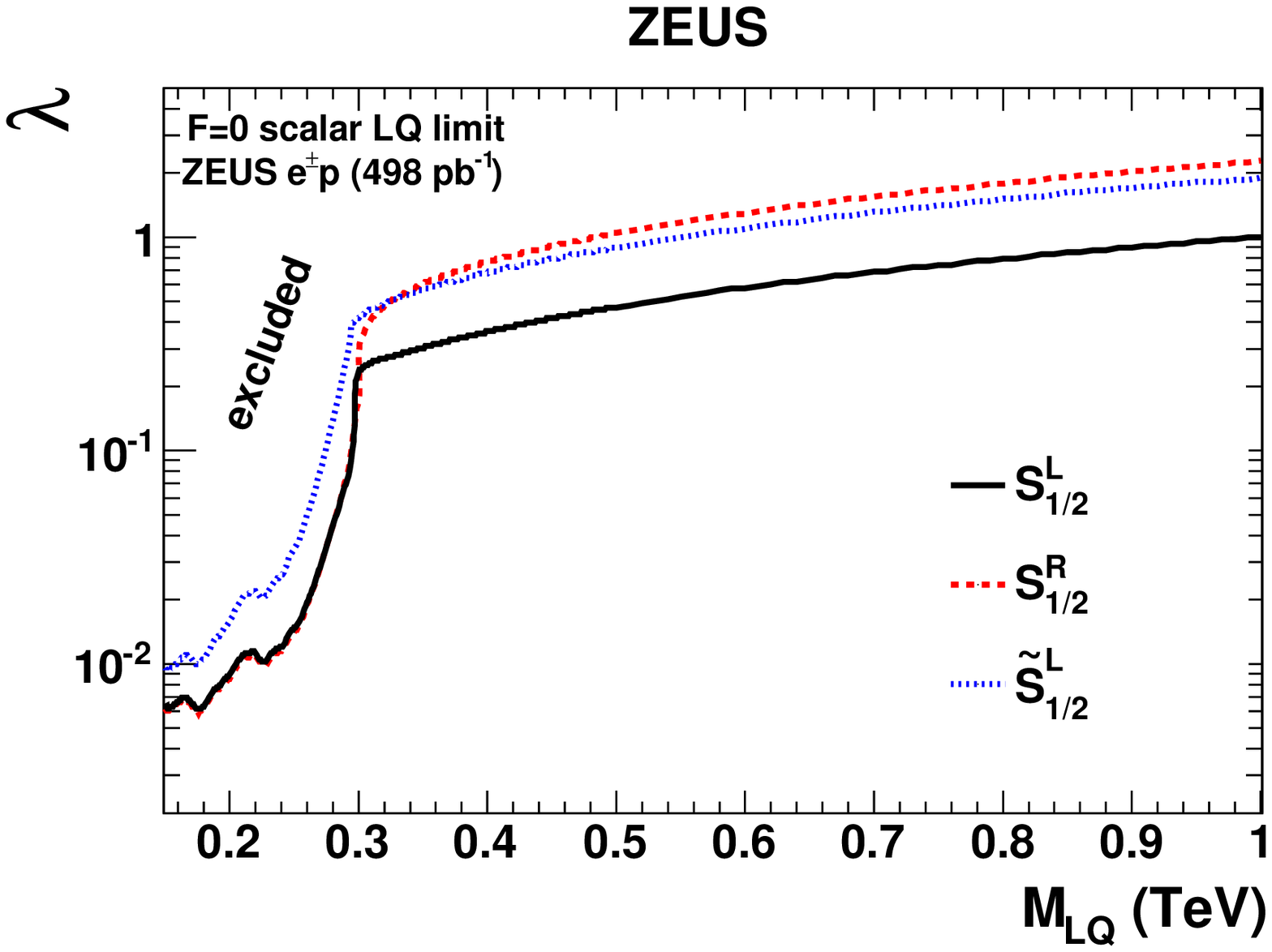}
\end{center}
\caption{
Coupling limits, $\lambda_\mathrm{limit}$, as a function of LQ mass for 
scalar F=0 BRW LQs. The areas above the curves are excluded according to Eq.~(\ref{eq:lam_lim}).}
\label{llimitS0}
\vfill
\end{figure}

\begin{figure}[p]
\vfill
\begin{center}
\includegraphics[width=11cm]{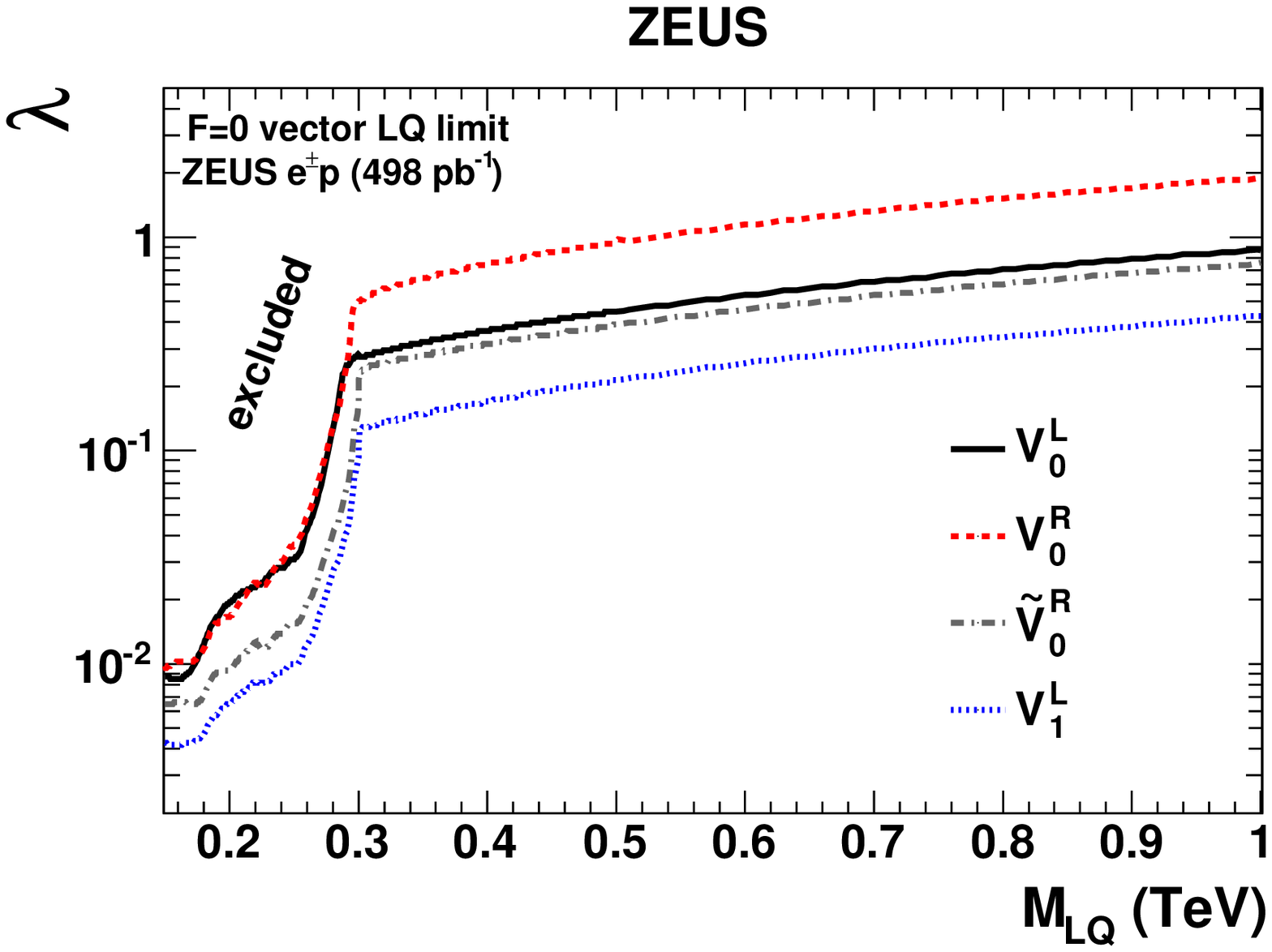}
\end{center}
\caption{
Coupling limits, $\lambda_\mathrm{limit}$, as a function of LQ mass for 
vector F=0 BRW LQs. The areas above the curves are excluded according to Eq.~(\ref{eq:lam_lim}).}
\label{llimitV0}
\vfill
\end{figure}

\begin{figure}[p]
\vfill
\begin{center}
\includegraphics[width=11cm]{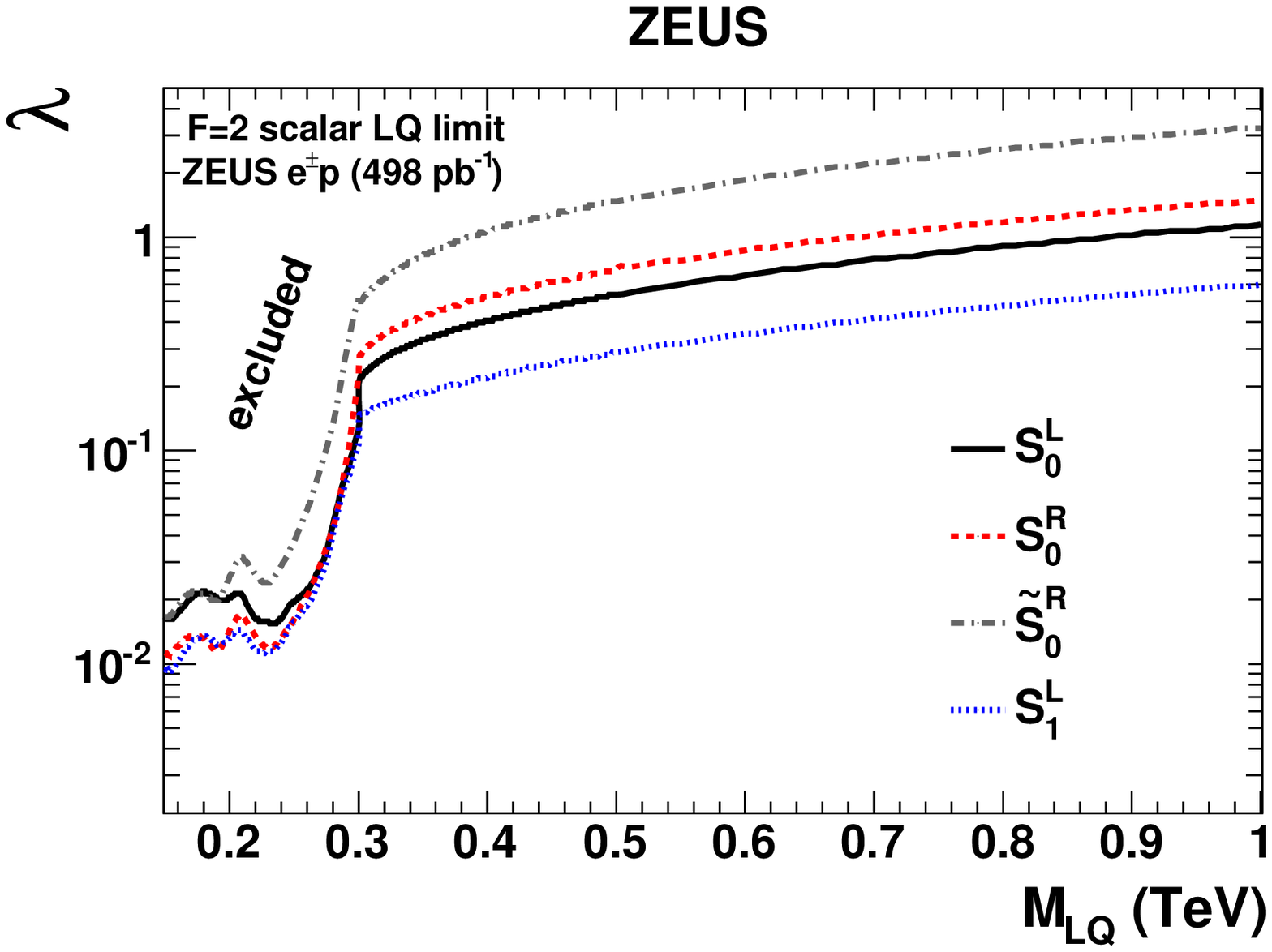}
\end{center}
\caption{
Coupling limits, $\lambda_\mathrm{limit}$, as a function of LQ mass for 
scalar F=2 BRW LQs. The areas above the curves are excluded according to Eq.~(\ref{eq:lam_lim}).}
\label{llimitS2}
\vfill
\end{figure}

\begin{figure}[p]
\vfill
\begin{center}
\includegraphics[width=11cm]{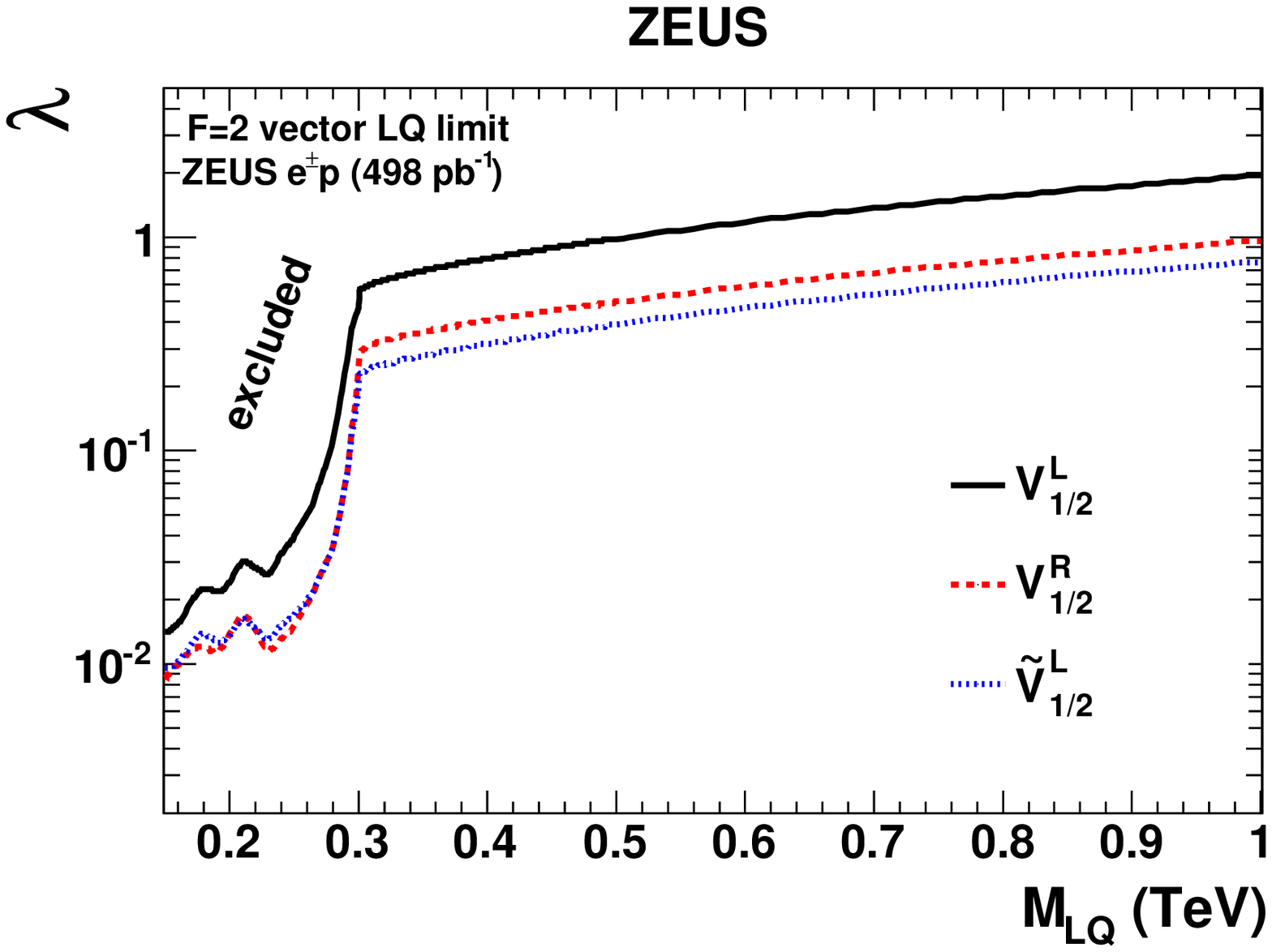}
\end{center}
\caption{
Coupling limits, $\lambda_\mathrm{limit}$, as a function of LQ mass for 
vector F=2 BRW LQs. The areas above the curves are excluded according to Eq.~(\ref{eq:lam_lim}).}
\label{llimitV2}
\vfill
\end{figure}

Figures~\ref{llimitS0}--\ref{llimitV2} show the coupling
limits on the scalar and vector LQs with $F=0$ and $F=2$,
respectively, where $F=3B+L$ is the fermion number of the LQ
and $B$ and $L$ are the baryon and lepton numbers, respectively.   The limits range from 0.004--0.017 for $M_\mathrm{LQ} = \unit[150]{\Gev}$, and from 0.43--3.24 for $M_\mathrm{LQ}=\unit[1]{\Tev}$. 
The lowest masses 
for which LQs with $\lambda=$~0.1 
and with $\lambda=$~0.3 are not excluded are summarised in Table~\ref{limit_table}. 
They range from 274 to \unit[300]{\Gev} for
$\lambda=$~0.1 and from 290 to \unit[699]{\Gev} 
for $\lambda=$~0.3.

The limit on the LQ state $\tilde{S}_{1/2}^L$ ($S_0^L$) can be interpreted as a limit on $\lambda \sqrt{\beta}$ for an up-type squark $\tilde{u}_L$ (a down-type squark $\tilde{d}_R$) in supersymmetric models with $R$-parity violation~\cite{np:b397:3}, where $\lambda$ is the coupling of $\tilde{u}_L$ to $eq$ ($\tilde{d}_R$ to $eq$ and $\nu q$) and $\beta$ is the branching fraction of the squarks to lepton ($e$ or $\nu$) and quark\footnote{The branching fractions of the squarks to $eq$ and $\nu q $ are assumed to be $\beta_{eq}=\beta,\ \beta_{\nu q}=0$ for $\tilde{u}_L$, and $\beta_{eq}=0.5 \, \beta,\ \beta_{\nu q}=0.5 \, \beta$ for $\tilde{d}_R$.}.

\begin{figure}[p]
\vfill
\begin{center}
\includegraphics[width=11cm]{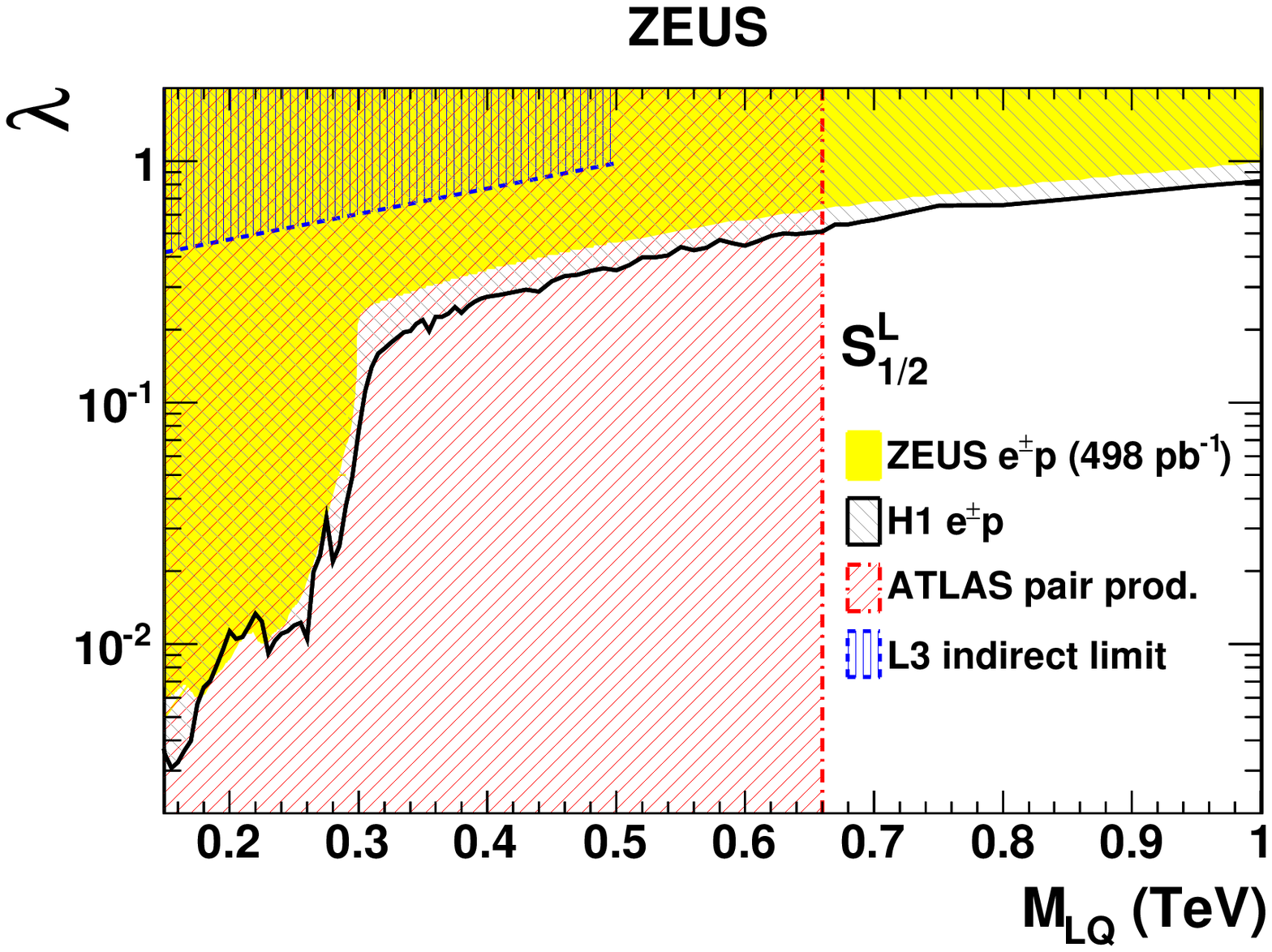}
\end{center}
\caption{
Coupling limits as a function of LQ mass for 
the $S_{1/2}^L$ LQ from ATLAS, L3, H1 and ZEUS.}
\label{llimitcomp1}
\end{figure}

\begin{figure}[p]
\vfill
\begin{center}
\includegraphics[width=11cm]{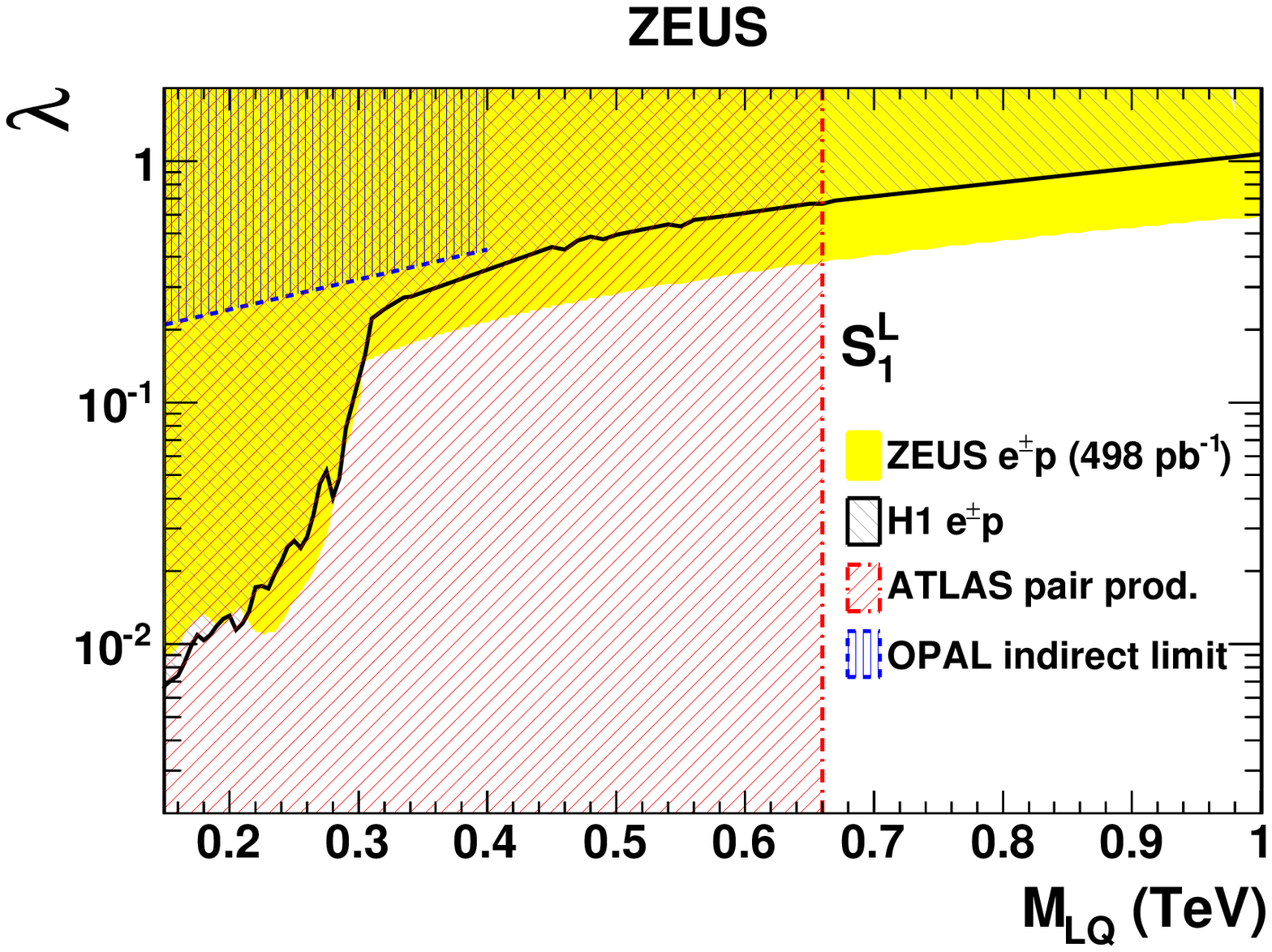}
\end{center}
\caption{
Coupling limits as a function of LQ mass for 
the $S_1^L$ LQ from ATLAS, OPAL, H1 and ZEUS.}
\label{llimitcomp2}
\end{figure}

Figure~\ref{llimitcomp1} (\ref{llimitcomp2}) shows the limits on the $S_{1/2}^L$ ($S_1^L$) LQ compared to the limits from ATLAS~\cite{Aad:2011ch}, H1~\cite{Collaboration:2011qaa} L3~\cite{Acciarri:2000uh} and OPAL~\cite{Abbiendi:1998ea}. The L3 and OPAL limits were evaluated up to \unit[500]{\Gev}. Limits using $pp$ or $p\bar{p}$ collisions are obtained from leptoquark pair production, which is independent of $\lambda$. Therefore it is not obvious whether the limits should be compared directly. The ATLAS exclusion range given in Fig.~\ref{llimitcomp2} also depends on the assumption that the branching ratio of the LQ state to electron and quark is one. Limits using $e^+ e^-$ collisions are obtained from a search for indirect effects in the process $e^+ e^- \rightarrow q \bar{q}$. In general, the limits from this analysis are significantly better than the LEP limits for $M_\mathrm{LQ} < \sqrt{s}$, and comparable for $M_\mathrm{LQ} > \sqrt{s}$. The limits obtained by ZEUS are similar to those obtained by H1.

\section{Conclusions}

Data recorded by the ZEUS experiment at HERA
were used to search for the presence of first-generation scalar and vector leptoquarks. 
The data samples include \unit[185]{$\pbi$} of $e^-p$ and \unit[181]{$\pbi$} of $e^+p$ collisions with polarised electrons and positrons. No resonances or other deviations from the Standard Model were found. The inclusion of data with unpolarised beams yields a total set of data corresponding to \unit[498]{$\pbi$}, which was used to set upper limits on the Yukawa coupling $\lambda$ for the 14 Buchm\"uller-R\"uckl-Wyler leptoquark states as a function of the leptoquark mass. Assuming $\lambda=0.3$, the mass limits range from 290 to \unit[699]{\Gev}. 
\vfill\eject

\section*{Acknowledgements}
\label{sec-ack}

\Zacknowledge

\vfill\eject

{
\ifzeusbst
  \bibliographystyle{./BiBTeX/bst/l4z_default}
\fi
\ifzdrftbst
  \bibliographystyle{./BiBTeX/bst/l4z_draft}
\fi
\ifzbstepj
  \bibliographystyle{./BiBTeX/user/l4z_epj-ICB}
\fi
\ifzbstnp
  \bibliographystyle{./BiBTeX/bst/l4z_np}
\fi
\ifzbstpl
  \bibliographystyle{./BiBTeX/bst/l4z_pl}
\fi
{\raggedright
\bibliography{./BiBTeX/user/syn.bib,%
              ./BiBTeX/user/myref.bib,%
              ./BiBTeX/bib/l4z_zeus.bib,%
              ./BiBTeX/bib/l4z_h1.bib,%
              ./BiBTeX/bib/l4z_articles.bib,%
              ./BiBTeX/bib/l4z_books.bib,%
              ./BiBTeX/bib/l4z_conferences.bib,%
              ./BiBTeX/bib/l4z_misc.bib,%
              ./BiBTeX/bib/l4z_preprints.bib}}
}
\vfill\eject

\begin{table}[ptb]
\begin{center}
\begin{tabular}
[c]{|c|c|c|c|c|}
\hline
 period  & lepton &  luminosity ($\mathrm{pb^{-1}}$) & $\langle P_e \rangle$ & $\sqrt{s}$ (GeV) \\ 
 \hline \hline
04--06 & $e^-$ & 106 & --0.27 & 318 \\ \hline
04--06 & $e^-$ & ~79 & ~\,0.30 & 318 \\ \hline
03--04 &  & ~17 && \\
06--07 & \raisebox{1.5ex}[-1.5ex]{$e^+$} & ~60 & \raisebox{1.5ex}[-1.5ex]{--0.37} & \raisebox{1.5ex}[-1.5ex]{318} \\ \hline
03--04 &  & ~21 && \\
06--07 & \raisebox{1.5ex}[-1.5ex]{$e^+$} & ~83 & \raisebox{1.5ex}[-1.5ex]{~\,0.32} & \raisebox{1.5ex}[-1.5ex]{318} \\ \hline
\hline
94--97 & $e^+$  & ~49 & 0 & 300 \\ \hline 
98--99 & $e^-$  & ~17 & 0 & 318 \\ \hline 
99--00 & $e^+$  & ~66 & 0 & 318 \\ \hline 
\end{tabular}
\end{center}
\caption{Details, including longitudinal polarisation, $P_e$, of the different 
data samples used.}
\label{tab:lumi}
\end{table}

\begin{table}[hbpt]
\begin{displaymath}
\begin{tabular}{c|crlcc}

LQ species & charge & production & decay & branching ratio & coupling
 \\ \hline \hline
F=0 &&&&& \\ \hline
$S_{1/2}^L$ & 5/3 & $e^+_R u_R$ & $e^+ u$ & 1 & $\lambda_L$ \\ \hline
$S_{1/2}^R$ & 5/3 & $e^+_L u_L $ & $e^+u$ & 1 & $\lambda_R$ \\ 
& 2/3 & $e^+_L d_L $ & $e^+ d$ & 1 & $-\lambda_R$ \\ \hline
${\tilde S}_{1/2}^L$ & -2/3 & $e^+_R d_R $ & $e^+d$ & 1 & $\lambda_L$ \\ \hline
$V_0^L$ & 2/3 & $e^+_R d_L $ & $e^+ d$ & 1/2 & $\lambda_L$ \\ 
&  & 
$$
& $\bar{\nu}_e u$ & 1/2 & $\lambda_L$\\ \hline
$V_0^R$ & 2/3 & $e^+_L d_R $ & $e^+ d$ & 1 & $\lambda_R$ \\ \hline
${\tilde V}_0^R$ & 5/3 & $e^+_L u_R $ & $e^+u$ & 1 & $\lambda_R$ \\ \hline
$V_1^L$ & 5/3 & $e^+_Ru_L $ & $e^+ u$ & 1 & $\sqrt{2}\lambda_L$\\ 
& 2/3 & $e^+_R d_L $ & $e^+ d$ & 1/2 & $-\lambda_L$\\ 
&  & 
$$
& $\bar{\nu}_e u$ & 1/2 & $\lambda_L$ \\ \hline \hline
F=2 &&&&& \\ \hline
$S^L_0$          &--1/3& $e^-_Lu_L$   & $e^-u$  &1/2& $\lambda_L$\\
                 &    & $        $   &$\nu_ed$ &1/2& $-\lambda_L$\\ \hline
$S^R_0$          &--1/3& $e^-_Ru_R$   & $e^-u$  &1  & $\lambda_R$\\ \hline
$\tilde{S}^R_0$  &--4/3& $e^-_Rd_R$   & $e^-d$     & 1 & $\lambda_R$\\ \hline
$S^L_1$          &--1/3& $e^-_Lu_L$   & $e^-u$     &1/2& $-\lambda_L$\\
                 &    &            &$\nu_ed$ &1/2& $-\lambda_L$\\
                 &--4/3& $e^-_Ld_L$   &$e^-d$& 1 &$-\sqrt{2}\lambda_L$ \\ \hline
$V^L_{1/2}$      &--4/3& $e^-_Ld_R$   &$e^-d$      & 1 & $\lambda_L$ \\ \hline
$V^R_{1/2}$      &--4/3& $e^-_Rd_L$   &$e^-d$     & 1 & $\lambda_R$ \\
                 &--1/3& $e^-_Ru_L$   &$e^-u$     & 1 & $\lambda_R$ \\ \hline
$\tilde{V}^L_{1/2}$&--1/3&$e^-_Lu_R$  &$e^-u$     & 1 & $\lambda_L$ 
\\ \hline
\end{tabular}
\end{displaymath}
\caption{Leptoquark species for fermion number $F=0$ 
and $F=2$  
as defined in the Buchm$\ddot{{u}}$ller-R$\ddot{{u}}$ckl-Wyler
model \protect\cite{pl:b191:442}
and the corresponding couplings. Those LQs that couple only to
neutrinos and quarks and therefore could not be produced at HERA
are not listed.
The LQ species are classified according to their spin ($S$
for scalar and $V$ for vector), their chirality ($L$ or $R$) and 
their weak
isospin ($0,1/2,1$). The leptoquarks ${\tilde S}$ and ${\tilde V}$ 
differ by
two units of hypercharge from $S$ and $V$, respectively. In addition,
 the
electric charge of the leptoquarks, the production channel,
 as well as
their allowed decay channels assuming lepton-flavour conservation
are displayed.
The nomenclature follows the Aachen convention
\protect\cite{zfp:c46:679}.}
\label{tab:brwlq}
\end{table}

\begin{table}[p]
\begin{center}
\begin{tabular}{|c||c|c|c|c|c|c|c|}
\hline
{ {LQ type (F=0)}}  &  $V_0^L$      & $V_0^R$
         &      $\tilde{V}^R_0$   &  $V_1^L$  &
           $S_{1/2}^L$  & $S_{1/2}^R$ & $\tilde{S}_{1/2}^L$  \\
\hline
$\lambda_{\mathrm{limit}}$ & 0.87 & 1.91 & 0.76 & 0.43 & 1.00 & 2.29 & 1.91 \\
\hline
{LQ type (F=2)}  & $S_0^L$     & $S_0^R$
         & $\tilde{S}^R_0$   &  $S_1^L$  &
           $V_{1/2}^L$  & $V_{1/2}^R$ & $\tilde{V}_{1/2}^L$  \\
\hline
$\lambda_{\mathrm{limit}}$ & 1.15 & 1.48 & 3.24 & 0.60 & 1.95 & 0.95 & 0.76 \\ \hline

\end{tabular}
\caption{Upper limit on the Yukawa coupling, $\lambda_\mathrm{limit}$ as defined in Eq.~(\ref{eq:lam_lim}), for the 14 BRW LQs at $M_\mathrm{LQ}=\unit[1]{\Tev}$.}
\label{limit_table_1TeV}
\end{center}
\end{table}

\begin{table}[p]
\begin{center}
\begin{tabular}{|c||c|c|c|c|c|c|c|}
\hline
{ {\bf LQ type (F=0)}}  &  $V_0^L$      & $V_0^R$
         &      $\tilde{V}^R_0$   &  $V_1^L$  &
           $S_{1/2}^L$  & $S_{1/2}^R$ & $\tilde{S}_{1/2}^L$  \\
\hline
$M_\mathrm{LQ}$(GeV) ($\lambda_\mathrm{limit}=0.1$) & 276 & 275 & 295 & 300 & 295 & 294 & 274 \\ \hline
$M_\mathrm{LQ}$(GeV) ($\lambda_\mathrm{limit}=0.3$) & 325 & 292 & 376 & 699 & 345 & 300 & 292 \\ 
\hline\hline
{\bf LQ type (F=2)}  & $S_0^L$     & $S_0^R$
         & $\tilde{S}^R_0$   &  $S_1^L$  &
           $V_{1/2}^L$  & $V_{1/2}^R$ & $\tilde{V}_{1/2}^L$  \\
\hline
$M_\mathrm{LQ}$(GeV) ($\lambda_\mathrm{limit}=0.1$) & 295 & 292 & 274 & 298 & 278 & 293 & 293 \\ \hline
$M_\mathrm{LQ}$(GeV) ($\lambda_\mathrm{limit}=0.3$) & 332 & 304 & 290 & 506 & 292 & 303 & 376 \\ \hline

\end{tabular}
\caption{Lower limit for the masses of the 14 BRW LQs for  
$\lambda_\mathrm{limit}$=0.1 and $\lambda_\mathrm{limit}$=0.3 as deduced from Eq.~(\ref{eq:lam_lim}).}
\label{limit_table}
\end{center}
\end{table}


%
%
\end{document}